\documentclass[aps, prd, preprintnumbers, floatfix, showpacs, showkeys, nofootinbib, 10pt]{revtex4-1}

\usepackage[dvips]{graphics}
\usepackage{amsmath,amsfonts,bm}
\usepackage{amssymb}
\usepackage{amsthm}
\usepackage{epsfig,bm}
\usepackage{feynmp}
\usepackage{graphicx,comment}
\usepackage{dcolumn,color}
\usepackage{mathtools}
\usepackage{fancyhdr}
\usepackage{bm}
\usepackage{cancel}
\usepackage{slashed}
\usepackage{verbatim}
\usepackage{listings}
\usepackage{graphicx,latexsym}
\usepackage{rotating}
\usepackage{color}
\usepackage{float}
\usepackage{enumerate}
\usepackage{array}
\usepackage{tabularx}
\usepackage{longtable}
\usepackage{booktabs}

 \def\cB{{\cal B}}  
   
   \def\cL{{\cal L}}
   
 \def\cR{{\cal R}}  
   
 \def\cR{{\cal R}}
\def\tr{\mathop{\rm Tr}}  \def\Tr{\mathop{\rm Tr}}  
\def\Re{\mathop{\rm Re}}     
   \def\GeV{{\rm GeV}}       
                 \def\d{\partial}

\def\beq{\begin{equation}}
\def\eeq{\end{equation}}
\def\bea{\begin{eqnarray}}
\def\eea{\end{eqnarray}}

\newcommand{\Aa}{\mathcal{A}}

\newcommand{\be}{\begin{equation}}
\newcommand{\ee}{\end{equation}}
\newcommand{\bear}{\begin{eqnarray}}
\newcommand{\eear}{\end{eqnarray}}
\newcommand{\ba}{\begin{array}}
\newcommand{\ea}{\end{array}}

\newcommand{\of}[1]{\!\left(#1\right)}

\newcommand{\avg}[1]{\left<#1\right>}
\newcommand{\vv}[1]{\mathbf{#1}}

\newcommand{\RET}{\nonumber \\}
\renewcommand{\And}{\quad\text{ and }\quad}

\begin{document}
\preprint{NYU-7-31-2016}
\preprint{\today}


\title{Central Production of $\eta$ via Double Pomeron Exchange and Double Reggeon Exchange in the Sakai-Sugimoto Model}


\author{Neil~Anderson$^1$, Sophia~K.~Domokos$^2$, Nelia~Mann$^3$}
\affiliation{$^1$ University of Maryland, College Park, MD 20742, USA~,}
\affiliation{$^2$ New York City College of Technology, Brooklyn, NY 11201, USA~,}
\affiliation{$^3$ Union College, Schenectady NY, 12308, USA}


\begin{abstract}
We use holographic QCD to study Pomeron- and Reggeon-mediated central production in Regge regime hadronic scattering. We focus specifically  on $\eta$ production in proton-proton collisions. While previous work studied the Pomeron-mediated process, the
mesonic Regge trajectories (``Reggeons'') we now incorporate contribute significantly at experimentally probed energies. We use the five-point open string amplitude (in flat space) to construct approximate propagators for the Reggeon states, and the five-point closed string amplitude for the Pomeron. Using these ``holographic'' Reggeons and Pomerons, and low-energy couplings derived in the Sakai-Sugimoto model, we compute the differential and total cross sections at $\sqrt{s}=29.1\GeV$. Our calculation of the total cross section is $\sigma =236 \ \mathrm{nb}$, while the experimentally measured value is $\sigma = 3859 \ \mathrm{nb}$.  The discrepancy between our result and the measured value may indicate that our model systematically underestimates the values of the relevant coupling constants.
\end{abstract}




\maketitle


\section{Introduction}

The holographic approach to hadronic physics is built  on the conjecture that there exists a \textit{duality} (or mapping) between non-supersymmetric gauge theories such as QCD, and string theory on a hyperbolically curved background.  AdS/CFT ---or, generically, gauge-string duality--- is a weak-strong mapping: perturbative calculations in the curved space string theory yield first principles predictions for strongly coupled hadronic physics. This idea has been used to analyze light meson and glueball states with reasonable success \cite{Sakai:2004cn, Brower:2000rp}, and has also been extended to study baryons \cite{baryons}.

Most work in holographic QCD assumes vanishing string length ($\alpha'\rightarrow 0$). In this approximation, the curved space string theory reduces to supergravity, a (relatively) tractable classical field theory with a finite number of fields.  Since string theory excitations are either massless or have masses proportional to the inverse string length, the $\alpha'\rightarrow 0$ limit truncates the string spectrum to massless fields of spin $2$ or less. This is certainly a reasonable approximation for very light states and very low energies. However, empirical evidence suggests that there are finite mass hadronic excitations which are {\it not} captured by this approximation \cite{PDG}. These include higher spin states such as the $\rho_3$ and $\rho_5$, as well as some of the more esoteric spin 1 mesons like the $h_1/b_1$ \cite{h1b1}. Even at low energies, then, it is clear that supergravity cannot be the whole story.

The problem with the supergravity limit is most apparent, however, in the Regge regime of hadronic scattering. (That is: high center-of-mass energy $s$ and small momentum transfer $t$ processes.) In fact, the first string theories\cite{Veneziano} were developed as phenomenological models for Regge regime hadronic scattering -- because ordinary point particle theories did not do the job. 

The Regge regime's small momentum transfer ($t<<0$) implies that the incoming states exchange strongly coupled objects (i.e. mesons or glueballs).  At the same time, the amplitude for exchanging a spin $J$ state scales as $s^J$, so higher spin excitations dominate for large center of mass energy $s$. Summing up the contributions from all $t$-channel poles requires an excursion into the complex $J$ plane, and yields an amplitude that is proportional to $s^{\alpha(t)}$, where $\alpha(t)$ is a linear function \cite{earlyRegge}. One of the great revelations of this program was the fact that this linear function is precisely the same as the one that describes the Regge trajectories of the mesonic spectrum.  String amplitudes in flat space exhibit precisely this kind of Regge-regime behavior.

Outside the Regge regime,  of course, flat space string amplitudes fail to capture the behavior of hadronic scattering. This -- and the advent of QCD -- led to the abandonment of string theory as a viable description of hadrons. While this is certainly true of flat space string theories, gauge-string duality suggests that  a {\it curved space} string theory -- the holographic dual of QCD -- might do the trick.

The Regge regime is thus essential to understanding (and testing) gauge-string duality. It is a setting in which the holographic approach  produces results that are distinct from phenomenological setups such as the chiral Lagrangian. It also opens the door to a wealth of experimental data that can shed light on parameters appearing in {\it low-energy} processes. 

Regge regime scattering has received less attention in the holographic QCD community because an honest treatment of the string dual requires prohibitively complicated curved space string calculations. Progress in this area thus requires some form of approximation.  For instance, the nice work of  \cite{bpstetc} studied $2\rightarrow 2$ and $2\rightarrow 3$ string scattering in a simplified spacetime ($AdS_5$), and using a perturbative expansion in the inverse AdS radius. This helped to elucidate the nature of the Pomeron trajectory (discussed in more detail below) in a holographic context, and made a connection between the ``hard'' (perturbative) Pomeron and the ``soft'' Pomeron of the Regge regime.

Others have used a more phenomenological approach, aiming to apply gauge-string techniques to actual experimental data.  The practical approach of \cite{DHM}, which we adopt here, is as follows. (1) We assume that string scattering amplitudes in weakly curved backgrounds have roughly the same {\it structure} as flat space amplitudes, but that the defining parameters (e.g Regge slopes and intercepts) take on different values than in flat space. (2)  Furthermore, the Regge regime amplitudes are determined in terms of low energy couplings calculable in the supergravity limit. This is a sensible approximation: Regge trajectories of mesons and glueballs couple in the same way as the lightest state on the trajectory (with corrections appearing at order $t/s$). We make predictions for Regge regime scattering processes by studying the exchange of the lightest mesons or glueballs, and ``Reggeizing'' the result: in other words, replacing single particle propagators with the Regge regime propagators we derive from string scattering.

This approach led to promising results for high center-of-mass energy $pp$ and $p\bar{p}$ scattering where Pomeron exchange dominates \cite{DHM}. In gauge-string duality, this is equivalent to the external protons exchanging a single closed string. Though distinct from standard Regge phenomenology in a variety of ways, this string-inspired analysis yielded a similar form for the scattering amplitude. In \cite{ADHM}, a $2\rightarrow 3$ scattering process was studied, in the hopes of illuminating the distinguishing characteristics of the gauge-string approach. The focus was on the $pp\rightarrow pp\eta$ and $pp\rightarrow pp\eta'$ processes, mediated by double Pomeron exchange. The five-point closed string amplitude has more complex structures than the four-point amplitudes of \cite{DHM}. In particular, the relatively small mass of the $\eta$ implied a  kinematic regime with an amplitude clearly distinct from that of naive Regge phenomenology, where central interaction vertices---not just propagators---are modified in the Regge regime. The $pp\rightarrow pp\eta$ process has an additional advantage: it is natural-parity violating, so that in the dual model the coupling arises from a Chern-Simons action. This means, in practice, that its interaction with Pomeron or Reggeon trajectories is dictated by a small number of couplings, whose values are relatively model independent (as described in more detail below). It can thus be used as a means to study gauge-string duality in general, rather than just one specific low energy model.

In this work, we complete the study of the $pp\rightarrow pp\eta$ process in the Regge regime.  The work of \cite{ADHM} included only Pomeron exchange. Here, we include Reggeon exchange as well, which gives a significant contribution at  the energy ($\sqrt{s}=29.1~$ GeV) where experimental data exists.  This process requires us to generalize the techniques of \cite{ADHM} to include open string processes. New subtleties arise in defining the Reggeziation prescription for open string amplitudes, and the modification of internal open string vertices.

The outline of this work is as follows: in Section \ref{SSreview} we briefly review holographic QCD and the Sakai-Sugimoto model. We then sketch the derivation of and results for the relevant meson, glueball, and baryonic vertices, most of which have already appeared in the literature.  Using these vertices, we compute the differential cross section for $pp\rightarrow pp\eta$ mediated by the lightest (and lowest spin) particles on the Regge trajectories, in section \ref{Feynman}. In Section \ref{Reggeization} we describe our new ``Reggeization'' procedure for lifting the low energy amplitudes to their Regge limit equivalents, and compute the differential cross section, including both Pomeron and Reggeon contributions, in the Regge regime. In Section \ref{Results} we display numerical results for the (differential) cross section and compare to existing data, finding a significant discrepancy between the two.  In Section \ref{Conclusion} we  summarize our results and provide concluding thoughts.  Appendix \ref{regge} includes the details of the calculations for Reggeization.

\section{\label{SSreview} Low Energy Couplings from Holographic QCD}

Holographic QCD is based on the assumption that there exists a string theory on a (hyperbolically) curved 5d space that is {\it dual} to QCD. In other words, the physics of the 4d gauge theory (QCD) can be mapped onto objects and processes in the 5d dual. While the exact form of the 5d dual theory is difficult to write down, approximate frameworks capturing the  symmetry and symmetry-breaking patterns of QCD can help us gain insight into hadronic physics, and into gauge-string duality itself.

A variety of holographic QCD models  have appeared in the literature, but most have the same general structure.  Weak coupling and weak curvature in the string theory map to large $N_c$ and strong coupling in QCD. Perturbative calculations in the string theory thus model strongly coupled large $N_c$ QCD behavior. In general, holographic QCD models make use of string theory in the presence of intersecting stacks of D-branes. When the number of branes in the ``color'' stack is taken to infinity, we can treat the color stack entirely in terms of its supergravity background (in other words, how the stack deforms spacetime). Closed strings living in the background have no color or flavor indices\footnote{The number of colors in QCD, $N_c$, is related to the radius of curvature of the background, but this does not mean that the closed strings have color charge.} and are thus dual to glueball states. Open strings (coresponding to mesons) can only end on stacks of (probe) D-branes that reproduce the flavor group. They transform in the (anti)fundamental of the flavor group, and are thus associated with mesonic states. Baryons arise in a somewhat more complicated manner.  In large $N_c$ QCD, their masses are proportional to $N_c$. In holography, this behavior is associated with wrapped D-branes which are point-like on the flavor branes. They interact with the fields on the flavor brane: one can equivalently think of them, therefore, as solitons of the flavor brane action.

For concreteness, we use the Sakai-Sugimoto model \cite{Sakai:2004cn}, which we now describe in more detail. This top-down construction has the advantage (over successful bottom-up models such as \cite{Erlich:2005qh}) that it contains relatively few free parameters, and unambiguously prescribes all low-energy couplings via the supergravity action.

The supergravity background is generated by a stack of $N_c$ $D4$-branes ($N_c\rightarrow\infty$). Wrapping the $D4$-branes around a spacelike direction of finite radius \cite{Witten:1998zw} induces confinement in the gauge theory.  The spacetime, as a result, assumes a cigar-like geometry. In the limit of vanishing string length, the closed strings propagating on the curved background reduce to the supergravity fields: the metric, the dilaton, etc.  The graviton lives in 10d, but one can decompose it according to the transformation properties of its components under the 4d Lorentz group. In addition to scalar and spin 1 pieces, the graviton contains a part that transforms as a $2^{++}$ state in 4d. This corresponds to the spin 2 glueball, the lightest state on the Pomeron trajectory, which is the only glueball of interest to us here.

Adding stacks of $N_f$ $D8$-branes and $N_f$ $\overline{D8}$-branes to the $D4$ background models the $U(N_f)_L\times U(N_f)_R$ flavor symmetry. Open strings ending on these ``flavor branes'' are dual to mesons.  When $N_f \ll N_c$, we can treat the $D8$s as probes. That is, we neglect their backreaction on the $D4$-brane geometry, and simply find the energy-minimizing shape they assume in the curved background. For this ``cigar'' background geometry, the  $D8$ and $\overline {D8}$ stacks join together and  form a $U$-shape. This is a geometric realization of chiral symmetry breaking: near the boundary of the space (dual to high energies) the flavor configuration looks like two stacks of branes, symmetric under swaps of the $D8$s and $\overline{D8}$s. In terms of the field theory, this is symmetric under chiral symmetry transformations. Near the bottom of the space (the tip of the cigar), the branes fuse into a single stack. The $U(N_f)_L\times U(N_f)_R$ flavor symmetry is broken to $U(N_f)_V$. The fields living on these flavor branes correspond to meson states.  The only relevant field for us is the non-abelian gauge potential $A_M$.  This has parts that transform as vectors under the 4d Lorentz group, which correspond to vector mesons like the $\rho$ and $\omega$. The parts of $A_M$ that transform as scalars, meanwhile, correspond to towers of pseudoscalars: the $\eta$, the $\pi$s, etc.

As $N_c\rightarrow\infty$, the $D4$ branes deform the spacetime to a metric of the form
\begin{align}
\label{eq:SSbackground}
ds^2=\tilde{g}_{MN}dx^Mdx^N=\left( \frac{U}{R}\right)^{3/2}\Big( \eta_{\mu\nu}dx^\mu dx^\nu + f(U)d\tau^2 \Big)+\left( \frac{U}{R}\right)^{-3/2}\Big( f(U)^{-1}dU^2+U^2d\Omega_4^2\Big)\\
e^{\phi}=g_s\left( \frac{U}{R}\right)^{3/4}~,\hspace{.4in} F_4=dC_3=\frac{2\pi N_c}{V_4}\epsilon_4~,\hspace{.4in} f(U)\equiv 1-\frac{U_{KK}^3}{U^3}~. \nonumber
\end{align}
The $x^\mu$ (for $\mu=0,1,2,3$) denote flat ``field theory'' coordinates parallel to the $D4$ stack, while $x^4\equiv \tau$ represents the direction wrapped by the $D4$s, with  $\tau\sim\tau+2\pi M_{KK}^{-1}$. The radial direction perpendicular to the $D4$ stack is $U$.  To eliminate a possible conical singularity, the radial coordinate $U$ is bounded from below at a finite value $U\ge U_{KK}$.
 The remaining directions form an $S^4$ transverse to the branes with $d\Omega_4^2$ the associated metric and $\epsilon_4$ its volume form. The length scale $R$ describes the curvature of the spacetime.
We can express the Yang-Mills coupling $g_{YM}$ and the Kaluza-Klein mass $M_{KK}$ in terms of supergravity parameters $U_{KK}$, the string coupling $g_s$, and the string length $l_s$:
\begin{align}
M_{KK}=\frac{3U_{KK}^{1/2}}{2R^{3/2}}~,\hspace{.4in} g_{YM}^2=2\pi M_{KK}g_sl_s~,\hspace{.4in} R^3=\pi g_sN_cl_s^3=\frac{g_{YM}^2N_cl_s^2}{2M_{KK}}~.
\end{align}
The low energy field theory will then depend only on these two free parameters, whose values we need to fix in order to generate a fully predictive framework.  We will take the standard approach of using the experimental values of the $\rho$ meson mass and the pion decay constant to do this.  It is possible that better agreement between this model and QCD might be obtainable by using a range of experimental values, but since we are interested only in a heuristic estimate for our final results (and because we know that Sakai-Sugimoto is not an exact QCD dual) it seems reasonable to use this simpler approach.  In the Sakai-Sugimoto model \cite{Sakai:2004cn}, one finds
\begin{align}
m_\rho=0.67 M_{KK}\hspace{.4in}\text{and}\hspace{.4in} f_\pi^2=\frac{1}{54\pi^4}g_{YM}^2N_c^2M_{KK}^2~.
\end{align}
The observed values are $m_\rho=776$ MeV, $f_\pi=93$ MeV. 

As noted above, the flavor $D8$s assume a $U$-shaped solution in the $(U,\tau)$ plane, described by
\beq
\tau (U) = \pm U_0^4 \sqrt{f(U_0)}\int\limits_{U_0}^U\frac{dU}{\left(\frac{U}{R}\right)^{3/2} f(U)\sqrt{U^8 f(U)-U_0^8 f(U_0)}}~.
\eeq
This solution depends on a constant parameter $U_0$, which denotes the lowest radial point the flavor branes reach along the $U$ direction. We set $U_0=U_{KK}$ in what follows. The lowest (radial) point of the flavor brane stack thus passes through the lowest point in the spacetime. In this case, the endpoints of the $D8$ stack as $U\rightarrow\infty$ lie at antipodal points on the $\tau$ circle.

The full system is described by the bulk supergravity and flavor brane actions:
\begin{align}
\label{eq:fullSS}
S&=S_{grav}+S_{DBI}+S_{RR}~,\\
\label{eq:gravaction} S_{grav}&=\frac{1}{2\kappa_{10}^2}\int d^{10}x~ \sqrt{-g}\left\{ e^{-2\phi}\left( R+4(\nabla\phi)^2\right)-\frac{(2\pi)^4l_s^2}{2\cdot 4!}F_4^2 \right\},\\
\label{eq:fullDBI}
S_{DBI}&=-T_8\int d^9x e^{-\phi} \tr\sqrt{-\det\left(\tilde{g}_{MN} + 2\pi\alpha' F_{MN} \right)},\\
S_{RR}&=T_8\int_{D8} C\wedge \Tr\left[\exp\left\{ \frac{F}{2\pi} \right\}\right]\sqrt{\hat{A}(\cR)}~.
\end{align}
Here $\kappa_{10}$ is the 10-dimensional Newton constant, $T_8=(2\pi)^{-8}l_s^{-9}$ is the $D8$-brane tension, $\tilde{g}_{MN}$ is the pullback of the background metric onto the $D8$-branes, and $F_{MN}=\d_M A_N-\d_N A_M-i[A_M,A_N]$ denotes the field strength of $U(N_f)$-valued gauge fields living on the branes.  The traces run over $U(N_f)$ indices, with normalization $\tr (T^aT^b)=\delta^{ab}/2$.  The $D8$-branes wrap the $S^4$ transverse to the $D4$-branes in \eqref{eq:SSbackground}, but we will assume that none of the fields depend on these coordinates. We can thus trivially integrate out the $S^4$ in what follows.

$S_{DBI}$ is the standard DBI D-brane action. When decomposed order by order in fields,  it yields kinetic and interaction terms for the brane gauge fields.

In the Ramond-Ramond (RR) action, meanwhile, $C$ denotes a sum over background RR forms. In our case, the only non-zero contribution comes from $C_3$, whose field strength is proportional to the volume form on the $S^4$. Since the integral is over the $D8$ worldvolume, we have to fill in the six remaining directions in the wedge product with factors of the gravitational curvature two-form $\cR$, and the gauge field strength two-form $F$. These factors will appear in the expansion of the exponential
\begin{align}
e^{F/2\pi}= 1 + \frac{1}{2\pi}F + \frac{1 }{2! (2\pi)^2} F\wedge F + \frac{ 1 }{3! (2\pi)^3} F\wedge F \wedge F +\dots
\end{align}
and of the
A-roof genus $\hat{A}(\cR)$,  a sum over Pontryajin classes ($p_i$):
\begin{align}
\hat{A}(\cR)= 1-\frac{1}{24}p_1(\cR)+\dots=1+\frac{1}{192\pi^2}\Tr \cR\wedge \cR+\dots
\end{align}
Note that this trace, unlike the one in $S_{RR}$, is over the Lorentz indices of the curvature two-form, related to the Riemann tensor as $\cR^{MN} = \frac{1}{2}R_{AB}^{\phantom{AB}MN}dx^A\wedge dx^B $. We stress that the central vertices in our $2\rightarrow 3$ process will come entirely from these Chern-Simons terms, and are related to the gauge and mixed gauge-gravitational anomalies in QCD.  As a result, the couplings and the low energy amplitude are relatively model-independent.

In principle, the above actions contain all of the possible excitations in the system, including baryons. As noted above, baryons are solitons of the flavor brane fields, but treating them as classical solitons is not only analytically difficult, it ignores
the fermionic nature of these objects. It will be more convenient and accurate, then, to introduce an effective action for the protons, that appropriately models their fermionic properties. We describe this effective action, $S_f$, in subsection \ref{fermiquad}.

Let us now consider the relevant parts of the spectrum and couplings in more detail.  Each brane and bulk field can be decomposed in a Kaluza-Klein tower of modes along the $U$ direction. The free equation of motion for the field amounts to an eigenvalue equation as a function of $U$, with the eigenvalue being the 4d momentum squared, $k^2$. Solutions to the equations of motion which vanish rapidly as $U\rightarrow\infty$ are ``normalizable modes'' and serve as the KK basis in which to expand the 10d fields. Individual meson and glueball states are dual to normalizable modes associated with a given eigenvalue -- i.e. a given fixed value of the 4d mass squared.  The cigar shape of the geometry (i.e. confinement) gives rise to a discrete spectrum. Each 10d field generates a tower of states, with the same spin, C, P quantum numbers, but higher and higher masses. The masses are inversely proportional to the radius of compactification in the $\tau$ direction.\footnote{The are actually two different types of KK towers. The first is the spectrum of normalizable modes, which is discrete due to the confinement induced by the cigar-shaped geometry. There is also a KK decomposition on the $x^4$ circle. The latter is not relevant to the present treatment, as it generates hadronic excitations suppressed in the Regge regime. We ignore it in what follows. One might also worry about the fermion states appearing in supergravity action. By imposing anti-periodic boundary conditions along the $\tau$ direction, however, these also become heavy and do not appear in this analysis.} Evaluating the supergravity action on these mode expansions, and integrating out the $U$ direction, one finds an effective 4d Lagrangian with an infinite number of 4d mesons and glueballs, along with couplings between them. Here, we are only interested in the lightest of these KK states, equivalent to the lightest states on the {\it leading} meson and glueball Regge trajectories -- which dominate in the Regge regime.

\subsection{Glueballs}

The 10d graviton $h_{MN}$ contains pieces that transform as spin 0, spin 1, and spin 2 objects in 4d. These are dual to spin 0, 1, and 2 glueballs. Only the latter ($h_{\mu\nu}$) is relevant to our analysis. The glueball spectrum was originally derived in \cite{Brower:2000rp}. Here we simply rewrite  \cite{Brower:2000rp}'s results in terms of our coordinates and parameters.

We expand the action \eqref{eq:gravaction} to second order in small perturbations $\tilde{h}_{AB}$ around the background metric $\tilde{g}_{AB}$ given in  \eqref{eq:SSbackground}:
\beq
g_{AB} = \tilde{g}_{AB} - \tilde{h}_{AB}~.
\eeq
Choosing a gauge where  $h$ is traceless we can write the spin 2 piece as an expansion over eigenfunctions along $u$, a dimensionless version of the radial coordinate $U$:
\beq
u\equiv \frac{U}{U_{KK}}~,
\eeq
so then
\beq
\tilde{h}_{\mu\nu}(x, U) =\left(\frac{U_{KK}}{R}\right)^{3/2}u^{3/2}\sum\limits_{n=1}^\infty \tilde{T}_n(u)h^{(n)}_{\mu\nu}(x)~,
\eeq
where $\tilde{T}_n(u)$ are the wavefunctions in the $u$ direction, and $h^{(n)}_{\mu\nu}$ are the coefficients in the expansion (which correspond to 4d states). Note that in terms of $u$, the function $f(u)$ appearing in the metric is $f(u)=1-u^{-3}$. Since we are only interested in the lightest of these spin 2 glueball states, we drop the $(n)$ indices and take
$\tilde{T}(u)\equiv \Tilde{T}_1(u)$.

Varying \eqref{eq:gravaction} with respect to $\tilde{h}_{\mu\nu}$, one finds an equation of motion that defines the radial wavefunction $\tilde{T}(u)$:
\beq
\partial_u\Big[u^4f \d_u \tilde{T}\Big] = -m_g^2\frac{R^3}{U_{KK}}u\tilde{T}=  -\frac{9}{4}\left(\frac{m_g}{M_{KK}}\right)^2 \ u\tilde{T} ~,
\eeq
where $m_g^2$ is the eigenvalue (on shell modes have $p^2=-m_g^2$). Imposing the boundary conditions that the wavefunction be normalizable as $u\rightarrow \infty$, and well-behaved at $u\rightarrow 1$ (which is $U \rightarrow U_{KK}$) restricts the possible values of $m_g^2$ to a discrete tower. These are the masses of the $2^{++}$ glueballs. The lightest mass state has $m_g^2= 1.57 M_{KK}^2$ \cite{Brower:2000rp}. Evaluating the action on the equations of motion yields
\begin{align}
S_{\mathrm{glueball}} = \left[\frac{g_s N_c^3 \ell_sM_{KK}^3}{3^5\pi}\int_{0}^{\infty} \ u \ \tilde{T}^2 \ du\right] \int d^4x \, \left[\d^\rho h^{\mu\nu}\d_\nu h_{\mu\rho} - \frac{1}{2}\d_\rho h^{\mu\nu}\d^\rho h_{\mu\nu} - \frac{m_g^2}{2} h^{\mu\nu} h_{\mu\nu}\right]~.
\end{align}
Choosing the normalization condition for $\tilde{T}(u)$ to be
\beq
1 = \frac{2g_s N_c^3 \ell_s M_{KK}^3}{3^5\pi}\int_{0}^{\infty} u \ \tilde{T}^2 \ du \eeq
we can integrate over $u$ to find the canonically normalized 4d kinetic terms in the action for the massive spin 2 glueball state:
\beq
S^{(2)}_{\mathrm{glueball}} = \frac{1}{2}\int d^4x \, \left[\d^\rho h^{\mu\nu}\d_\nu h_{\mu\rho} - \frac{1}{2}\d_\rho h^{\mu\nu}\d^\rho h_{\mu\nu} - \frac{m_g^2}{2} h^{\mu\nu} h_{\mu\nu}\right]~.
\eeq

\subsection{Mesons}

The meson spectrum and wavefunctions are determined by expanding the DBI action to quadratic order in the brane gauge field $A_M$.  The $u$-dependent wavefunctions for these states were determined and discussed in \cite{Sakai:2004cn}. We review the essential results here.

$A_M$ contains modes which transform as 4d spin 1 and spin 0 states, dual to (axial) vector mesons and  pseudoscalars, respectively.
As we did for the glueballs, we can expand solutions to the spin 1 and spin 0 states' equations of motion in the basis of $u$-dependent wavefunctions on the $D8$ branes.  Again, the boundary conditions produce a discrete spectrum. The corresponding wavefunctions alternate parity under exchange of $\overline{D8}$ branes to $D8$s, which corresponds to switching $U(N_f)_L\leftrightarrow U(N_f)_R$ in the dual field theory.  The lowest modes (the ones of interest to us) have even wavefunctions and correspond to vector states such as the $\rho$ and $\omega$, with the next modes up having odd wavefunctions and corresponding to axial vector states\footnote{For this work we are only interested in the neutral version of the $\rho$-meson, which for notational simplicity we will refer to as $\rho$.}.

The quark mass is zero in the Sakai-Sugimoto model, so the pseudoscalars, which include $\pi$ and $\eta$, are the Goldstone bosons of chiral symmetry breaking. Their duals are housed in the longitudinal part of $A_M$ --- which makes sense because $A_M$ is dual to the (broken) axial current. Our gauge choice for the $A_M$ solution essentially determines whether the pseudoscalars are carried in the $A_U$ or in the longitudinal part of $A_\mu$, $\d^\mu A_\mu$ (which are related by a gauge transformation). We will use $A_U=0$ gauge in what follows.  Note that the Sakai-Sugimoto framework implicitly assumes that the vector flavor symmetry is preserved, so all mesons of given spin and parity have the same mass (e.g. the $\rho$ and the $\omega$).

Expanding the DBI action \label{eq:fullDBI} to quadratic order in $A_\mu$ (again, assuming no dependence on the $S^4$), we have
\begin{equation}
S_{DBI} = - \frac{g_{YM}^2N_c^2}{108\pi^3}\int_1^\infty  d^4x \  du \tr\left[ \frac{3}{4\sqrt{u  f(u)}}\eta^{\mu\nu}\eta^{\rho\sigma}F_{\mu\rho}F_{\nu\sigma}
+\frac{2}{3}M_{KK}^2\sqrt{uf(u)}\eta^{\mu\nu}F_{\mu u}F_{\nu u}+\dots\right]~.
\end{equation}
Note that $u$ only parameterizes half of the  $D8$-$\overline{D8}$ stack, so we impose the even(odd)-ness of the wavefunctions as a boundary condition at the bottom of the space. In the UV (as $u\rightarrow \infty$), the solutions must decay quickly enough to guarantee a canonically normalized kinetic term in the 4d action. Overall, we have
\begin{align}\label{Asol}
A_\mu(x,u)&=\sum\limits_{n=1} A_\mu^{(n)}(x)\psi_n(u) + \d_\mu\phi(x)\psi_0(u)~,
\end{align}
where $\phi(x)$ represents the massless Goldstone bosons ($\eta$ and $\pi$, e.g.), and $A_\mu^{(1)}$ corresponds to $\rho$ and $\omega$ mesons. The  $\psi_n$'s obey the equation of motion and orthogonality relation
\begin{align}
-\frac{4}{9} \sqrt{u  f(u)} \ \d_u\left(\sqrt{u^2 f(u)}   \ \d_u\psi_m \right)=-\frac{m_n^2}{M_{KK}^2}\frac{1}{\sqrt{u f(u)}}\psi_n\quad\text{and}\quad \frac{N_c^2 g_{YM}^2}{36\pi^3}\int du
\frac{1}{ \sqrt{u f(u)}  }  \psi_n\psi_m=\delta_{mn}~.
\end{align}
These expressions hold for all of the flavor generators, so that $\rho$ and $\omega$ have the same wavefunction.

The mode $\phi$ is massless, and its wavefunction $\psi_0$ is
\begin{align}
\psi_0(u) =  4\pi\sqrt{3N_c l_sM_{KK}}g_{YM}^2 \int_1^u du'\frac{1}{u'\sqrt{u'^3-1}}
\end{align}
This mode corresponds to the pseudoscalar mesons. Everywhere we take $N_f=3$.  We are ultimately interested in an $\eta$ final state, which is a linear combination of the $T^0$ and $T^8$ components:\footnote{The $\pi$ state also plays a role in the process we are studying, since it is included in solitonic configurations that are dual to protons.  However, we will treat the proton using an effective fermion action, so the only part of the pseudoscalar action we are interested in here pertains to the $\eta$ particle.}
\begin{align}
\eta = \cos\theta_m \eta^8 -\sin\theta_m\eta^0~.
\end{align}
 $\theta_m=-11.05^o$ is the (experimentally observed) mixing angle. Note that both the $\eta^0$ and $\eta^8$ appear in our analysis: the former couples to glueballs in the central vertex, while both couple to the vector mesons.

Integrating out the $u$ direction, one finds the quadratic action for the mesonic states of interest,
\begin{align}
S^{(2)}_{\mathrm{meson}} &=\int d^4 x \left[-\frac{1}{4}f^{(\rho)}_{\mu\nu}f^{(\rho)\mu\nu}-\frac{1}{2}m_\rho^2\rho_\mu \rho^{\mu}-\frac{1}{4}f^{(\omega )}_{\mu\nu}f^{(\omega)\mu\nu}-\frac{1}{2}m_\omega^2\omega_\mu \omega^{\mu}\right.\nonumber\\
&\qquad\qquad\qquad-\left.\frac{1}{2}\d_\mu\eta\d^\mu\eta-\frac{1}{2}m_{\eta}^2\eta^2\right]~,
\end{align}
where we have defined
\beq
f^{(\rho)}_{\mu\nu}\equiv \d_\mu\rho_\mu-\d_\mu\rho_\nu~,
\eeq
and similarly for $\omega$.

\subsection{Protons}\label{fermiquad}

In AdS/QCD models, baryons are dual to finite volume D-branes.  In the Sakai-Sugimoto model specifically, they are $D4$-branes embedded in the worldvolume of the $D8$-branes, wrapping the $S^4$. In terms of the worldvolume fields, these wrapped branes are 5d, curved space solitons  for which no analytic solutions exist.  The solutions have been studied in various limits \cite{baryons}, and a full (numerical) solution was found in \cite{Bolognesi:2013nja}.

All of these classical treatments are flawed, however, in that they treat baryons as a spinless states. Spin arises only with the semi-classical quantization of the collective coordinates on the soliton.  It is helpful, then, to introduce an ``effective'' fermion field $\cB$ on the  D8-brane worldvolume, with a $u$-depended effective mass \cite{Hong:2007kx, Hong:2007ay, Hong:2007dq, Park:2008sp, Domokos:2010ma}. This is somewhat ad-hoc because this field does not appear in the DBI action, but it does more accurately model the Lorentz transformation properties of the proton states (and hopefully the proton-meson and proton-glueball couplings). Many of the parameters appearing in the action are derived by assuming that the baryon is a 5d instanton. The effective fermion technique is therefore an educated guess at the semi-classical quantization of the baryon states.

We use this effective fermion approach in what follows.  Our starting point is the 5d action of \cite{Hong:2007ay}, which encodes both the effective fermion's kinetic terms and its coupling to the graviton \cite{Domokos:2010ma} and gauge fields:
\beq
S = \int d^4x \, dw \, \left[-i\bar{\mathcal{B}}\gamma^MD_M\mathcal{B} - im_b(w)\bar{\mathcal{B}}\mathcal{B} + g_5(w)\frac{\rho_{b}^2}{e^2(w)}\bar{\mathcal{B}}\gamma^{MN}F_{MN}\mathcal{B} \right]
\eeq
Note that our conventions differ slightly from those of \cite{Hong:2007ay}.  The indices $M,N$ run over the 5 dimensions $\{0, 1, 2, 3, w\}$, where $w$ is a (dimensionful) redefinition of the dimensionless radial coordinate $u$,
\beq
w = \frac{3}{2M_{KK}}\int_{1}^{u} \frac{du'}{\sqrt{u'^3 f(u')}}~,
\eeq
chosen so that the radial coordinate has finite range $[-w_{\mathrm{max}}, w_{\mathrm{max}}]$, with
\begin{align}
w_{\mathrm{max}} = \frac{3}{2 M_{KK}}\int_{1}^{\infty} \frac{du}{\sqrt{u^3 f(u)}} =  \frac{3}{M_{KK}} \, \frac{\sqrt{\pi}\Gamma[\frac{7}{6}]}{\Gamma[\frac{2}{3}]}~.
\end{align}

The 5d vector field $A_M$ transforms in the adjoint of the full $N_f=3$ flavor brane gauge group. Since we are interested in protons alone, however, we can just look at the $U(2)$ corresponding to isospin in the field theory:
\beq
A_M = \frac{1}{2}\hat{A}_M\mathbb{I}_2 + \frac{1}{2}A^a_m\sigma^a
\eeq
where $\sigma^a$ are the sigma matrices, and $\mathbb{I}_2$ is the $2 \times 2$ identity matrix. $\hat{A}$ contains the $\omega$ meson, while $A^3$ contains the $\rho$.  The gauge covariant derivative $D_M$ is given by $D_M = \partial_M - iA_M$ as usual, and the field strength $F_{MN}$ is $ F_{MN}= \partial_MA_N - \partial_NA_M - i[A_M,A_N]$.

We have written the 5d action in terms of  effective $w$-dependent ``couplings'' $e(w), ~g_5(w)$ and ``mass'' $m_b(w)$. These are simply names for the various metric factors that appear as coefficients in 5d action.  $e(w)$ is the (inverse) coefficient of the gauge kinetic term,
\beq
\frac{1}{e^2(w)} =
\frac{\pi \, f_{\pi}^2}{2 \, M_{KK}} \, u(w)~.
\eeq

The baryon's ($w$-dependent) ``mass'' is then given by
\beq
m_b(w) =m_b^{(0)} \cdot u + m_{e}\qquad\text{with}\qquad m_b^{(0)}= \frac{\lambda N_c}{27\pi} \cdot M_{KK}\qquad\text{and}\qquad m_e\approx \frac{1}{3} m_b^{(0)} (M_{KK}\rho_b )^2
\eeq
where the first term comes from wrapping the D4-brane around the finite volume $S^4$, and the second comes from the Coulomb energy of the baryon.  $\rho_b$ represents the size of the soliton, 
\beq
\rho_b^2=\sqrt{\frac{3}{10}} \ \frac{N_c}{\pi^3 f_\pi^2}~.
\eeq

The 5d fermion $\mathcal{B}$ is a four-component spinor, and a two-component (fundamental) representation of $U(2)$ isospin. The first component of the isospin fundamental corresponds to the proton, the second to the neutron.

The quadratic order terms in the fermion action are
\beq
S_f \supset \int d^4x \, dw \, \Big[-i\bar{\mathcal{B}}\gamma^M\partial_M\mathcal{B} - im_b(w)\bar{\mathcal{B}}\mathcal{B}\Big]
\eeq
We can split $\mathcal{B}$ into two spinors $\mathcal{B}_{L, R}$, where $\gamma^5\mathcal{B}_{L, R} = \pm \mathcal{B}_{L, R}$.  Assuming, separation of variables we can write these as an expansion in terms of $w$-direction eigenfunctions,
\beq
\mathcal{B}(x^\mu, w) =\sum\limits_{n=1} \left(\begin{array}{c} f_{Ln}(w)B^{(n)}_L(x^\mu) \\ f_{Rn}(w)B^{(n)}_R(x^\mu)\end{array}\right)~.
\eeq
As before, we are only interested in the first state on the KK tower, and drop the $(n)$ index in what follows.

The 5d equation of motion for the fermion,
\beq
\gamma^m\partial_m\mathcal{B} + m_b(w)\mathcal{B} = 0~,
\eeq
becomes (after some simplification)
\begin{align}f_R(w)\bar{\sigma}^{\mu} \partial_\mu B_R(x^\mu) + \partial_w f_L(w)B_L(x^\mu) + m_b(w)f_L(w)B_L(x^\mu) = 0 \\
f_L(w) \sigma^{\mu} \partial_\mu B_L(x^\mu) - \partial_w f_R(w) B_R(x^\mu) + m_b(w) f_R(w)B_R(x^\mu) = 0 \, .
\end{align}
If we want to reassemble the 4d Weyl spinors $B_{L,R}$ into a single 4d Dirac spinor $B$, we need
\begin{align}
\partial_w f_L(w) + m_b(w) f_L(w) = m_B f_R(w), \hspace{.75in} -\partial_w f_R(w) + m_b(w)f_R(w) = m_B f_L(w)
\end{align}
where $m_B$ is a 4d mass.

By analyzing the equations of motion, we can show that $f_L(w) = \pm f_R(-w)$.  One can then solve (numerically) for the $f_L(w)$ eigenfunctions, which must obey the normalization condition
\beq
1 = \int_{-w_{\mathrm{max}}}^{w_{\mathrm{max}}} |f_L(w)|^2 \, dw = \int_{-w_{\mathrm{max}}}^{w_{\mathrm{max}}} |f_R(w)|^2 \, dw \, ,
\eeq
in order to guarantee a canonically normalized 4d kinetic term.
The 4d action for the proton is then just the usual action for a massive Dirac fermion,
\beq
S^{\text{quadratic}}_f = \int d^4x \Big[-i\bar{\Psi}\gamma^\mu\partial_\mu \Psi - im_B\bar{\Psi}\Psi\Big]~.
\eeq
Here $\Psi$ stands for the isospin component of the spinor $B$ that corresponds to the proton.

\subsection{Low energy couplings}\label{lowenergycouplings}

The couplings involving mesons and glueballs are calculated in the literature \cite{Sakai:2004cn,Domokos:2009cq, ADHM}.  Most of the proton-proton-meson \cite{Hong:2007ay,Hong:2007dq,Park:2008sp} and -glueball couplings \cite{DHM} are also known.  We just sketch the 10d origins of these couplings here.

\subsubsection{Meson-meson-meson and glueball-glueball-meson vertices}

The central vertices in the $pp\rightarrow pp\eta$ process of interest couple two glueballs to the $\eta$, or two vector mesons to the $\eta$. These natural parity-violating couplings come from the Ramond-Ramond action, $S_{RR}$. Expanding in the gravitational curvature $\cR$ and  gauge field strength $F$, we find
\begin{align}\label{SRRexp}
S_{RR}&=\int_{D8} C_3\wedge\left[  \frac{1}{768\pi^3}\Tr(F)\wedge\Tr(\cR\wedge \cR ) + \frac{1}{48\pi^3} \Tr(F\wedge F\wedge F)\right]+\dots\nonumber\\
&=\int_{D8} dC_3\wedge\left[ \frac{1}{768\pi^3} \Tr(A)\wedge\Tr(\cR\wedge \cR) + \frac{1}{48\pi^3} \omega_5(A)\right]+\dots~,
\end{align}
where $\omega_5(A)$ is the Chern-Simons five-form.  Again, we assume no fluctuations along the $S^4$. The background RR four-form $F_4=dC_3$ is (crucially) proportional to the volume form on the $S^4$, so integrating out the $S^4$ just gives a factor proportional to the volume of a unit $S^4$. The first term  in equation \eqref{SRRexp} contains the glueball-glueball-$\eta$ coupling, and was studied in \cite{ADHM}. The second term consists of gauge fields alone, and produces meson-meson-$\eta$ couplings derived in \cite{Sakai:2004cn,Domokos:2009cq}, which generate the Reggeon-Reggeon-$\eta$ vertex. While naively one might think that there also glueball-meson-eta couplings, it is straightforward to see that they must vanish: the coupling will involve a factor of the epsilon tensor, and there is no way to contract this into the Lorentz indices of the glueball and meson states in a way that yields a non-zero result. This will simplify our calculations significantly in later sections.

The 4d effective couplings are integrals over $u$ of products of wavefunctions. For instance, writing the gauge field as a sum of even and odd parity parts, $V$ and $A$, the gauge Chern-Simons piece gives
\begin{align}
S_{RR}\supset& \frac{N_c}{24\pi^2}\int_{\mathcal{M}_5} \tr [A \wedge dA \wedge dA + 3 A \wedge dV \wedge dV]\\
=& -\frac{N_c}{12\pi^2}\int_{\mathcal{M}_5} \epsilon^{\mu\nu\rho\sigma} \, \tr[A_{\mu} \, \partial_u A_{\nu} \, \partial_\rho A_\sigma + A_{\mu} \, \partial_{\nu}A_{\rho} \, \partial_uA_{\sigma} \ + \ 3A_{\mu} \, \partial_uV_{\nu} \, \partial_{\rho}V_{\sigma} + 3A_{\mu} \, \partial_{\nu}V_{\rho} \, \partial_zV_{\sigma}]
\end{align}
where in the final equality we use the fact that we are in $A_u=V_u=0$ gauge. Inserting wavefunction expansions\footnote{In principle these wavefunction expansions should sum over all KK states in the mass spectrum, but as we are only interested in the lightest states we drop the rest of the sum in what follows.} for $V= v_\mu(x)\psi_V (u)$, integrating by parts, and using the cyclicity of the trace, we find
\beq
S_{RR} \supset \frac{N_c}{4\pi^2}\int_{\mathcal{M}_5} \epsilon^{\mu\nu\rho\sigma} \, \tr[T^aT^bT^c]\Big\{(\psi_0'\psi_V^2) \, \partial_{\mu}\phi^a \, \partial_{\nu}v_{\rho}^{b} \, v_{\sigma}^{c}
\Big\}~.
\eeq
Integrating over the $u$ direction, we obtain (constant) 4d couplings, with action terms of the form
\beq
S_{RR} \supset -\int d^4 x \, \epsilon^{\mu\nu\rho\sigma}\eta\Big\{g_{\rho\rho\eta} \, \partial_{\nu}\rho_{\rho} \, \partial_\mu\rho_{\sigma} + g_{\omega\omega\eta} \, \partial_{\nu}\omega_{\rho} \, \partial_\mu\omega_{\sigma}  \Big\}
\eeq
The numerical values for all coupling constants are collected in table \ref{table}, in section \ref{Results}.

The (more complicated) glueball-glueball-$\eta$ vertices \cite{ADHM} are derived from the mixed gauge-gravitational Chern-Simons term in a similar way. The first term in the second line of the RR action \eqref{SRRexp} takes the form
\begin{align}
S_{RR}\supset\frac{N_c}{1536\pi^2}\int d^5x \tilde{\epsilon}^{MNPQR} \Tr(A_M)R_{NPST}R_{QR}^{\phantom{NP}TS}~.
\end{align}
Because the (gauge index) trace is over $A_M$ alone, only the $U(1)$ part of the D-brane gauge field couples to the gravitons (housed in the Riemann tensors). We are interested in $\eta$, which lives in the longitudinal part of $A_\mu$.
Plugging in the wavefunctions in the radial direction and integrating out we are left with interaction terms of the form
\beq
S_{RR} \supset \int d^4x \epsilon^{\mu\nu\rho\sigma} \, \eta \, \Big\{\hat{g}_{hh\eta}\partial_{\mu}h^{\alpha}_{\nu}\partial_{\sigma}h_{\rho\alpha} \ + \ \tilde{g}_{hh\eta}\partial_{\mu}\partial^{\alpha}h_{\nu}^{\beta}\partial_{\rho}\left(\partial_{\beta}h_{\sigma\alpha} - \partial_{\alpha}h_{\sigma\beta}\right)\Big\}~.
\eeq
As was noted in \cite{ADHM}, the structure of the above expression is dictated by the form of the Chern-Simons action.  This structure is therefore relatively model-independent, though the specific values of the coupling constants depend on the details of the Sakai-Sugimoto model.  The structure of this coupling  will play a significant role in the behavior of the Pomeron-mediated portion of the cross section for central production.

\subsubsection{\label{lambda} Proton-proton-meson and proton-proton-glueball vertices}

The proton-proton-meson and proton-proton-glueball couplings come from the fermion-fermion-gauge field and fermion-fermion-graviton terms in $S_f$. The former, described in detail in \cite{Hong:2007ay,Hong:2007dq}, are obtained by plugging the mode expansions of the fermions and gauge fields into $S_f$. 

There are two types of 5d interaction terms, referred to as the minimal coupling, and the magnetic coupling.
After integrating out the $S^4$, these become:
\beq\label{fermiint}
S_f \supset-\int d^4x \, dw \, \Big[\bar{\mathcal{B}}\gamma^M A_{M}\mathcal{B}+\frac{u(w)N_c}{\sqrt{30}M_{KK}}\bar{\mathcal{B}}\gamma^{MN} F_{MN}\mathcal{B}\Big]
\eeq
Again, $A_N$ refers to the 5d gauge field. As before, we work in $A_u=A_w=0$ gauge.

Selecting the lightest mass eigenfunction in $A_\mu$ and integrating over $w$, one finds the effective 4d minimal coupling of the lightest vector to the fermionic isospin doublet:
\beq
S_{V} \supset -\lambda^{(V)}\int d^4x \left[\bar{B}\gamma^{\mu}v_{\mu}B\right]~.
\eeq
where
\begin{align}
\lambda^{(V)} = \int dw \, |f_L(w)|^2\psi_{V}(w)~. \\
\end{align}
The  couplings from the second term in \eqref{fermiint},  $\left[\bar{\mathcal{B}}\gamma^{MN}F_{MN}\mathcal{B}\right]$, are derived in a similar way \cite{Hong:2007ay,Hong:2007dq}. Note that there are two types of couplings from this term: one that comes from the
$\gamma^{ 5\mu}F_{w\mu}=\gamma^{5\mu}\d_w A_\mu$ and one from $\gamma^{\mu\nu}F_{\mu\nu}$.
Integrating out the $w$ direction, one finds the effective 4d magnetic couplings to the lightest vector mesons:
\beq
S_V\supset \int d^4x \left[ \hat{\lambda}^{(V)} \bar{B} \gamma^\mu v_\mu B  + \tilde{\lambda}^{(V)} \bar{B} \gamma^{\mu\nu} f^{(V)}_{\mu\nu} B\right]
\eeq
where
\begin{align}
\hat{\lambda}^{(V)}&=-\frac{N_c}{M_{KK}}\sqrt{\frac{2}{15}}\int_{-w_{max}}^{w_max} dw u(w) |f_L(w)|^2\d_w\psi_V(w)\\
\tilde{\lambda}^{(V)}&=\frac{N_c}{M_{KK}^2}\sqrt{\frac{1}{30}}\int_{-w_{max}}^{w_max} dw f_L(w)f_R(w)\psi_{V}(w)~.
\end{align}

When we rewrite this in terms of couplings involving the physical proton, $\rho$, and $\omega$, we find that the magnetic couplings vanish for the $\omega$, because the instanton carries only non-abelian field strength.  For the $\rho$ meson, the effects of $\lambda^{(V)}$ and $\hat{\lambda}^{(V)}$ simply add together.  Thus we have action terms of the form
\beq
S_V \supset \int d^4 x \Big\{\lambda_{pp\rho}\bar{\Psi}\gamma^{\mu}\rho_{\mu}\Psi + \tilde{\lambda}_{pp\rho}\bar{\Psi}\gamma^{\mu\nu}f^{(\rho)}_{\mu\nu}\Psi + \lambda_{pp\omega}\bar{\Psi}\gamma^{\mu}{\omega}_{\mu}\Psi \Big\}~.
\eeq

The glueball coupling \cite{Domokos:2010ma} comes from  the metric factor  in the fermion kinetic term,
\begin{align}
S_f \supset \int d^4x \, dw \, i\bar{\mathcal{B}}(\tilde{g}_{\mu\nu}+\tilde{h}_{\mu\nu})\gamma^\mu\partial^\nu\mathcal{B}~.
\end{align}
This approximately gives a coupling of the graviton to the energy momentum tensor of the protons $T^{(\Psi)}$ at the bottom of the $D8$ stack. In this case, we have the effectively 4d interaction
\begin{align}
S_f \supset \lambda_{pph} \int d^4x \, h^{\mu\nu}T^{(\Psi)}_{\mu\nu}~.
\end{align}

\subsubsection{Form Factors}
In addition to the couplings, we also include proton form factors at the interaction vertices with the glueballs and mesons. The form factors associated with the vector mesons are commonly taken to have a standard dipole form,
\beq
A_{v}(t) = \frac{1}{\left(1 - \frac{t}{M_{dv}^2}\right)^2}
\eeq
with an empirically determined dipole mass $M_{dv}$ \cite{L.A.}. We follow that convention here.

There are no experimental data on glueball interactions with the proton. However, in the holographic dual theory, the spin 2 glueball couples to the proton's energy momentum tensor, so we can simply use the energy momentum tensor's form factor(s).  The matrix element for the energy momentum tensor between proton states can be written as
\beq
\langle p', s'|T_{\alpha\beta}|p, s \rangle = \bar{u}(p', s')\left[A_g(t) \, \frac{\gamma_{\alpha}P_{\beta} + \gamma_{\beta}P_{\alpha}}{2} + B_g(t) \, \frac{i(P_{\alpha}\sigma_{\beta\rho} + P_{\beta}\sigma_{\alpha\rho})k^{\rho}}{4m_p} + C_g(t) \, \frac{(k_{\alpha}k_{\beta} - \eta_{\alpha\beta}k^2}{m_p}\right]u(p, s)
\eeq
Here, $k = p - p'$ and $P = \frac{1}{2}(p + p')$.  As described in \cite{DHM}, only the first term contributes significantly to scattering processes in the Regge limit. We approximate this with a dipole form as well,
\beq
A_{g}(t) = \frac{1}{\left(1 - \frac{t}{M_{dg}^2}\right)^2}
\eeq
with the dipole mass $M_{dg}^2$ calculated from a skyrmion treatment of the protons \cite{Cebulla:2007ei}.  Both dipole masses are given in table \ref{table}, in section \ref{Results}.

\subsection{\label{summary} Summary of Full Low-Energy Lagrangian}
We can now assemble the low energy Lagrangian relevant to our model:
\begin{align}\label{Seff4d}
\cL = \cL^{(p)}_{0}+\cL^{(m)}_{0}+\cL^{(g)}_{0}+\cL^{(pm)}_{int}+\cL^{(mm)}_{int}+\cL^{(pg)}_{int}+\cL^{(mg)}_{int}~,
\end{align}
where the quadratic proton, meson, and glueball Lagrangians are given (in position space) by
\begin{align}
\cL^{(p)}_{0} &= i\bar{\Psi}\slashed{\d}\Psi-m_p\bar{\Psi}\Psi\\
\cL^{(m)}_{0}&=-\frac{1}{4}f^{(\rho)}_{\mu\nu}f^{(\rho)\mu\nu}-\frac{1}{2}m_\rho^2\rho_\mu \rho^{\mu}-
-\frac{1}{4}f^{(\omega )}_{\mu\nu}f^{(\omega)\mu\nu}-\frac{1}{2}m_\omega^2\omega_\mu \omega^{\mu}
-\frac{1}{2}\d_\mu\eta\d^\mu\eta-\frac{1}{2}m_{\eta}^2\eta^2\\
\cL_0^{(g)}&=\frac{1}{2} \left[\d^\rho h^{\mu\nu}\d_\nu h_{\mu\rho} - \frac{1}{2}\d_\rho h^{\mu\nu}\d^\rho h_{\mu\nu} - \frac{m_g^2}{2} h^{\mu\nu} h_{\mu\nu}\right]~,
\end{align}
and the interaction terms (in momentum space) are
\begin{align}
\cL^{(pm)}_{int}=&\left [-\lambda_{pp\rho}\bar{\Psi}\gamma^\mu\rho_{\mu}\Psi+i\tilde{\lambda}_{pp\rho}\bar{\Psi}\gamma^{\mu\nu}f^{(\rho)}_{\mu\nu}\Psi-\lambda_{pp\omega}\bar{\Psi}\gamma^\mu\omega_\mu\Psi  \right] A_v(t)\\
\cL^{(mm)}_{int}=& ~\epsilon^{\mu\nu\rho\sigma}\Bigg\{g_{\rho\rho\eta}\eta \, p_{\nu}\rho_{\rho}(p) \, p'_\mu\rho_{\sigma}(p') + g_{\omega\omega\eta}\eta \, p_{\nu}\omega_{\rho}(p) \, p'_\mu\omega_{\sigma}(p')  \Bigg\}\\
\cL^{(pg)}_{int}=&-ig_{ppg} h_{\mu\nu}\bar{\Psi}\Gamma_{(g)}^{\mu\nu}\Psi\\
\cL^{(mg)}_{int}=&~\left\{ \hat{g}_{hh\eta} \epsilon^{\mu\nu\rho\sigma}\eta\eta^{\alpha\beta}p_\mu h_{\nu\alpha}(p)p'_\sigma h_{\rho\beta}(p')+\tilde{g}_{hh\eta} \epsilon^{\mu\nu\rho\sigma}\eta\eta^{\alpha\beta}\eta^{\gamma\delta}p_\mu p_\beta h_{\nu\gamma}(p)p'_\rho\left( p'_\delta h_{\sigma\alpha}(p')-p'_\alpha h_{\sigma\delta}(p')\right)\right\}~,
\end{align}
with
\begin{align}
\Gamma_{(g)}^{\mu\rho} = \check\lambda_{pph}\left[\frac{A_g(t)}{2}\left(\gamma^{\mu}P^{\rho} + \gamma^{\rho}P^{\mu}\right)\right],
\end{align}
for $P_\mu = \frac{p+p'}{2}$ and $t=-(p-p')^2$.
The values of these couplings, calculated as described above, are summarized in Table \ref{table}.
\
\section{\label{Feynman} The Feynman Amplitude and the Cross Section}

Having determined the effective Lagrangian for the low energy $pp\rightarrow pp\eta$ process using the Sakai-Sugimoto model, we can now calculate the low energy cross section for $\eta$ central production mediated by the $2^{++}$ glueballs, and by the $\rho$ and $\omega$ mesons.  In the next sub-section we introduce the kinematics and define variables. After that, we write down the  scattering amplitudes and determine the cross section.

\subsection{Kinematics and Phase Space}
Let us first review the kinematics of $2\to 3$ scattering and approximations to be made in the Regge limit.  We denote the four-momenta of the incoming protons as $p_1$ and $p_2$, that of outgoing protons as $p_3$ and $p_4$, and that of $\eta$ as $p_5$. In  the ``mostly-plus'' convention, these satisfy the mass-shell conditions
\begin{equation}\label{eq:massshell}
    p_1^2 = p_2^2 = p_3^4 = p_4^4 = -m_p^2 \And p_5^2 = -m_{\eta}^2~,
\end{equation}
and energy/momentum conservation
\begin{align}\label{eq:consof4momentum}
    p_1 + p_2 = p_3 + p_4 + p_5 .
\end{align}
We will denote the momentum of the mediating mesons and glueballs as $k_1 = p_1 - p_3$ and $k_2 = p_2 - p_4$.

The Mandelstam variables that define the five-point amplitude can be written as
\begin{align}
    s = -(p_1+p_2)^2, \qquad t_1 = -(p_1-p_3)^2, \qquad t_2 = -(p_2-p_4)^2, \qquad s_1 = -(p_1-p_4)^2, \qquad s_2 = - (p_2-p_3)^2~.
\end{align}
For scattering in the CM frame, we take
\begin{align}
    p_1 = (E,0,0,p), \qquad p_2 = (E,0,0,-p),\qquad p_3 = (E_3,\vv{q}_3, p_{3z}),\qquad p_4 = (E_4,\vv{q}_4, p_{4z}),\qquad p_5 = (E_5,\vv{q}_5,p_{5z})~.
\end{align}
In order to effectively compare with experimental results, we will be parametrizing in terms of four quantities: $\{t_1, t_2, x_F, \theta_{34}\}$.  The third of these, $x_F$, is the fraction of longitudinal momentum, defined according to
\begin{align}
    x_F = -\frac{p_{5z}}{p} = \frac{p_{3z} + p_{4z}}{p} = x_1 - x_2 \qquad \text{with} \qquad p_{3z} = x_1 p, \qquad p_{4z} = -x_2p.
\end{align}
The fourth quantity, $\theta_{34}$, is the angle between the outgoing protons' {\it transverse} momenta:
\begin{align}
    \theta_{34} = \theta_{4} - \theta_{3},\qquad \vv{q}_3 = (q_3\cos\theta_3,q_3\sin\theta_3),\qquad \vv{q}_4 = (q_4\cos\theta_4,q_4\sin\theta_4) .
\end{align}

In what follows, we will consider a Regge limit such that 
\beq
s \gg s_1,s_2 \gg t_1,t_2,m_i^2
\eeq
where $m_i$ stands for any of the external particle masses, and the quantity
\begin{align}
    \mu \equiv \frac{s_1 s_2}{s}
\end{align}
is held fixed.\footnote{In much of the literature, the symbol $\eta$ is used for this quantity; our choice is made to avoid confusion with the $\eta$ meson.} As was shown in \cite{ADHM}, the cross section is in the Regge limit is dominated by the region of phase space near $x_F = 0$, allowing the the total cross section to be approximated as
\begin{align}\label{eq:crosssection}
    \sigma \approx \frac{1}{4(4\pi)^4s^2} \int \avg{|\Aa|^2}\,\ln\of{\frac{s}{\mu}}\,d\theta_{34}\,dt_1\,dt_2.
\end{align}
We may also use the Regge limit to approximate
\begin{align}
    \mu \approx m_5^2 - t_1 - t_2 + 2\sqrt{t_1t_2}\cos\theta_{34},\qquad s_1 \approx s_2 \approx \sqrt{s\mu},\qquad q_3 \approx \sqrt{-t_1}, \qquad q_4 \approx \sqrt{-t_2}~.
\end{align}

\subsection{Amplitude}

\begin{figure}
\vspace{.4in}
\begin{center}
\begin{fmffile}{graph1}
\begin{fmfgraph*}(2,1.3)
\fmfleft{i1,i2}
\fmfright{o1,o2,o3}
\fmf{fermion,label.side=right,label=$p_3$}{i1,v1}
\fmf{fermion,label.side=right,label=$p_4$}{v1,o1}
\fmf{fermion,label.side=left,label=$p_1$}{i2,v2}
\fmf{fermion,label.side=left,label=$p_2$}{v2,o3}
\fmffreeze
\fmf{wiggly,label=$D^{v}_{\mu\rho}$}{v2,v3}
\fmf{wiggly,label.side=left,label=$D^{v}_{\nu\sigma}$}{v1,v3}
\fmf{dashes}{v3,o2}
\fmfdot{v1,v2,v3}
\fmfv{label=$\Gamma_v^{\mu}$\hspace{.1in}}{v2}
\fmfv{label=$\Gamma_v^{\nu}$\hspace{.1in}}{v1}
\fmfv{label.angle=180,label=$V_v^{\rho\sigma}$}{v3}
\end{fmfgraph*}
\end{fmffile}
\begin{fmffile}{graph3}
\begin{fmfgraph*}(2,1.3)
\fmfleft{i1,i2}
\fmfright{o1,o2,o3}
\fmf{fermion,label.side=right,label=$p_3$}{i1,v1}
\fmf{fermion,label.side=right,label=$p_4$}{v1,o1}
\fmf{fermion,label.side=left,label=$p_1$}{i2,v2}
\fmf{fermion,label.side=left,label=$p_2$}{v2,o3}
\fmffreeze
\fmf{curly,label=$D^{g}_{\mu\rho\alpha\epsilon}$}{v2,v3}
\fmf{curly,label.side=left,label=$D^{g}_{\nu\sigma\beta\phi}$}{v1,v3}
\fmf{dashes}{v3,o2}
\fmfdot{v1,v2,v3}
\fmfv{label=$\Gamma_g^{\mu\rho}$\hspace{.1in}}{v2}
\fmfv{label=$\Gamma_g^{\nu\sigma}$\hspace{.1in}}{v1}
\fmfv{label.angle=180,label=$V_g^{\alpha\epsilon\beta\phi}$}{v3}
\end{fmfgraph*}
\end{fmffile}
\vspace{.4in}
\caption{The Feynman diagrams for double Reggeon exchange and double Pomeron exchange.}
\label{fig:diagrams}
\end{center}
\end{figure}
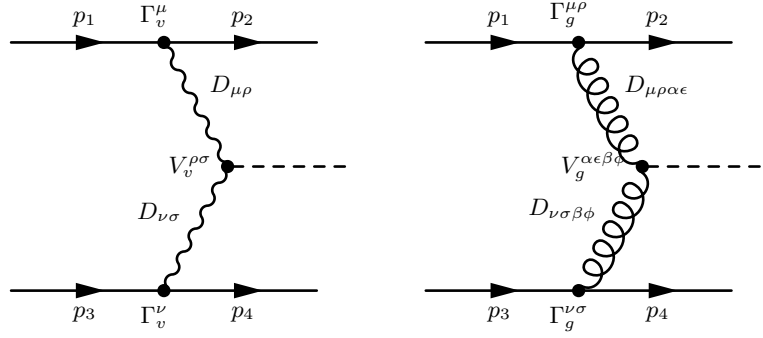

The $pp\rightarrow pp\eta$ amplitude contains terms corresponding to the exchanges of vector mesons (both $\rho$ and $\omega$) and of glueballs.\footnote{We could also include the exchange of axial vector mesons.  However, as will be discussed in section \ref{Reggeization}, these terms are suppressed in the Reggeization process, because the intercepts for their associated Regge trajectories are lower than those for the vector mesons and glueballs.}
\begin{equation}
\label{Asum}
    \mathcal{A} = \Aa_g + \sum\limits_{\rho,\omega}\Aa_v .
\end{equation}
The structures of the two vector-mediated terms are the same, only differing the the values of coupling constants and masses.  The Feynman diagrams for each are shown in Figure \ref{fig:diagrams}.

To write down this amplitude, we will need the following vertices and propagators. 
The propagator for the vector mesons is
\beq
D^v_{\mu\nu}(k) =\frac{i}{k^2+m_v^2}\left( \eta_{\mu\nu}+\frac{k_\mu k_\nu}{m_v^2}\right)~.
\eeq
where $k$ is the momentum of the vector meson, and $m_v$ is the mass of either $\rho$ or $\omega$.  Only the first term will contribute to the Regge limit amplitude.
The spin-2 glueball propagator, as given in \cite{prop} is
\begin{align}
D^g_{\mu\rho\nu\sigma}(k) = \frac{-i d_{\mu\nu\rho\sigma}}{k^2+m_g^2}
\end{align}
where
\begin{align}
d_{\mu\rho\nu\sigma}(k) &= \frac{1}{2}\left(\eta_{\mu\nu}\eta_{\rho\sigma} + \eta_{\mu\sigma}\eta_{\rho\nu}\right) - \frac{1}{2m_g^2}\left(k_{\mu}k_{\sigma}\eta_{\rho\nu} + k_{\mu}k_{\nu}\eta_{\rho\sigma} + k_{\rho}k_{\sigma}\eta_{\mu\nu} + k_{\rho}k_{\nu}\eta_{\mu\sigma}\right)\RET
&+ \frac{1}{24}\left[\left(\frac{k^2}{m_g^2}\right)^2 - 3\left(\frac{k^2}{m_g^2}\right) - 6\right]\eta_{\mu\rho}\eta_{\nu\sigma} - \frac{(k^2 - 3m_g^2)}{6m_g^4}\left(k_{\mu}k_{\rho}\eta_{\nu\sigma} + k_{\nu}k_{\sigma}\eta_{\mu\rho}\right) + \frac{2k_{\mu}k_{\rho}k_{\nu}k_{\sigma}}{3m_g^4},
\end{align}
Here, $k$ is the momentum of the glueball. Only the first term of this tensor will end up mattering in the Regge limit, just as above. 

The vector-vector-pseudoscalar vertex is given by
\begin{align}
    V^{\alpha\beta}_{(v)} = 2ig\epsilon^{\alpha\beta\gamma\delta}k_{1\gamma}k_{2\delta}
\end{align}
where $g$ is either $g_{\rho\rho\eta}$ or $g_{\omega\omega\eta}$, and
the glueball-glueball-pseudoscalar vertex is
\begin{align}
   V^{\alpha\epsilon\beta\phi}_{(g)} = 2i\epsilon^{\mu\epsilon\rho\beta}k_{1\mu}k_{2\rho}\Bigg\{\eta^{\alpha\phi}\Big(\hat{g}_{hh\eta} - \tilde{g}_{hh\eta}(k_1\cdot k_2)\Big) + \tilde{g}_{hh\eta}k_{1}^{\phi}k_{2}^{\alpha}\Bigg\}.
\end{align}
Finally, we have the proton-proton-vector vertex
\begin{align}
   \Gamma_v^{\mu} = i\Big(-\lambda\gamma^{\mu} + \tilde{\lambda}[\gamma^{\mu}, \gamma^{\nu}]k_{\nu}\Big)
\end{align}
where $\lambda$ can be either $\lambda_{pp\rho}$ or $\lambda_{pp\omega}$ and $\tilde{\lambda}$ is either $\tilde{\lambda}_{pp\rho}$ or $\tilde{\lambda}_{pp\omega} = 0$,
and the proton-proton-glueball vertex
\begin{align}
    \Gamma_g^{\mu\rho} = \frac{i\check{\lambda}A(t)}{2}\Big(\gamma^{\mu}P^{\rho} + \gamma^{\rho}P^{\mu}\Big) .
\end{align}
where $P \equiv (p+p')/2$ with $p = p_{1,2}$ and $p' = p_{3,4}$.

This leads to the amplitude terms
\beq
\label{Ag}
\Aa_g = \Big(\bar{u}_3\hat{\Gamma}^{\mu\rho}u_1\Big)D^g_{\mu\rho\alpha\epsilon}\hat{V}^{\alpha\epsilon\beta\phi}D^g_{\nu\sigma\beta\phi}\Big(\bar{u}_4\hat{\Gamma}^{\nu\sigma}u_2\Big)~,
\eeq
and
\beq
\label{Av}
\Aa_v = \Big(\bar{u}_3\Gamma^{\mu}u_1\Big)D^v_{\mu\alpha}V^{\alpha\beta}D^v_{\nu\beta}\Big(\bar{u}_4\Gamma^{\nu}u_2\Big)~.
\eeq

\subsection{The Differential Cross Section}

In order to calculate the differential cross section, we then want to find the square of the magnitude of the amplitude, average over incoming spin states, and sum over outgoing spin states, which gives:
\beq
\label{ave}
\avg{|\Aa|^2} = \frac{1}{4}\sum_{\text{spins}}|\Aa|^2
\eeq
In the Regge limit, using equations \eqref{Asum}, \eqref{Ag}, \eqref{Av}, and \eqref{ave}, the differential cross section for the low energy process becomes
\begin{align}
    \frac{d^3\sigma}{dt_1\,dt_2\,d\theta_{34}} = \frac{t_1t_2\sin^2\theta_{34}}{(4\pi)^4}\,\ln\of{\frac{s}{\mu}}
&\Bigg\{\frac{16\tilde{\lambda}^2_{pp\rho}g_{\rho\rho\eta}^2(4\tilde{\lambda}^2_{pp\rho}t_1t_2 - \lambda^2_{pp\rho}t_1 - \lambda^2_{pp\rho}t_2)}{(t_1 - m_{\rho}^2)^2(t_2 - m_{\rho}^2)^2}\RET
+\Bigg[\frac{\check{\lambda}_{pph}^2A(t_1)A(t_2)s\Big(\hat{g}_{hh\eta} - \tilde{g}_{hh\eta}\sqrt{t_1t_2}\cos\theta_{34}\Big)}{(t_1 - m_g^2)(t_2 - m_g^2)}
&- \frac{2\lambda_{pp\rho}^2g_{\rho\rho\eta}}{(t_1 - m_{\rho}^2)(t_2 - m_{\rho}^2)} - \frac{2\lambda_{pp\omega}^2g_{\omega\omega\eta}}{(t_1 - m_{\omega}^2)(t_2 - m_{\omega}^2)}\Bigg]^2
\Bigg\}~.
\end{align}

The overall factor of $t_1 t_2 \sin^2\theta_{34}$ in this expression comes directly from the natural parity violation: any vertex structure we could write down that violates natural parity would yield such dependence.  Furthermore,  the term associated with Pomeron exchange involves additional dependence on $\theta_{34}$, in a structure that derives from the Chern-Simons action (and is thus independent of the specifics of the Sakai-Sugimoto model). 

Note also  that the first term in the curly brackets comes from the magnetic coupling between the protons and the vector mesons; it applies only to the $\rho$ meson, and does not end up contributing to the cross term $\Aa_g^{*}\Aa_{v} + \Aa_{v}^{*}\Aa_g$.

\section{\label{Reggeization} Reggeizing the Propagators}

In the Regge regime, central production of the $\eta$ meson in $pp$ collisions should involve exchange of meson and glueball Regge trajectories. In this section we determine how to modify low energy amplitudes using the effective Lagrangian \eqref{Seff4d} to include entire Regge trajectories of possible mediators.

The magnitude of the contribution from a given Regge trajectory is primarily determined by the value of the Regge intercept.  If the Regge trajectory of the mediator is described at positive $t$ as $J = \alpha(m^2) = \alpha_0 + \alpha' m_J^2$, the amplitude will be proportional to $s^{\alpha(0)}$, which means the cross section scales as $s^{2(\alpha_0 - 1)}$.  At the highest energies, the Pomeron dominates: its intercept is $\alpha_{g0} = 1.08$ (based on fitting to proton-proton scattering data \cite{DHM}).  The intercept for vector mesons $\rho$ and $\omega$ is $\alpha_{v0} = 0.456$, from a linear fit to known meson masses \cite{PDG}. As the energy decreases, the contribution from these Reggeons increases.  At the energies we are most interested in, the Reggeons provide the dominant contribution.  The next largest contribution would come from the axial-vector meson trajectories corresponding to the $a_1$ and $f_1$ mesons.  However, the known experimental masses suggest\footnote{The available data here is much less reliable or comprehensive than for the vector mesons.} that the intercept for these trajectories is even lower, $\alpha_{a0} \sim -0.4$, and that as a result their contribution is negligible at the energies we are working at.  We have therefore omitted them from this work.


The trajectories described above are the effectively 4d trajectories one can (in principle) derive from the string dual system. We now turn to the properties of these states in the dual model. Mesonic Regge trajectories correspond to open string excitations, while glueball trajectories correspond to closed string excitations. Though these strings live in a 10d curved space, the spacetime curvature is weak. We will therefore assume that we can model curvature effect by simply shifting the 4d Regge trajectory parameters from their flat space values into agreement with physical mesons and glueballs.

Our goal is to use 5-string amplitudes (of both open and closed strings) to understand how to ``Reggeize'' propagators of light meson and glueball states, to take into account contributions from the whole trajectory.  We will start with calculations in flat-space bosonic string theory, and then modify these to account for the physical trajectories.  In previous work, this technique was applied to proton-proton scattering via Pomeron exchange, by analyzing the Virasoro-Shapiro amplitude \cite{DHM} and to central eta production via double Pomeron exchange, by studying the 5-closed-string amplitude \cite{ADHM}.  The procedure for proton-proton scattering via Reggeon exchange is based on a comparison of Veneziano's original result with the classic flat-space bosonic open string calculation \cite{Veneziano}.  Our results are also consistent with what can be obtained by considering the analyticity requirements for multiparticle scattering amplitudes \cite{OldBrower}.

We first briefly review the 4-string calculations in order to establish our procedure for generalizing flat-space bosonic strings to physical Regge trajectories.  We then summarize the previous results using the 5-closed-string amplitude.  Finally, we use our procedure on the 5-open-string amplitude.  In appendix \ref{regge} we present the details of all relevant calculations for this section.

\subsection{Review of Reggeization Procedure for Elastic Proton-Proton Scattering}

Suppose we were analyzing elastic proton-proton scattering via the exchange of Reggeons and Pomerons.  In order to Reggeize the propagators we should begin with the 4-open-string and 4-closed-string amplitudes in flat space bosonic string theory, before modifying these amplitudes to account for the physical Regge trajectories.  

For open strings, we want the exchanged states to have odd spin, so we should project out poles associated with the exchange of even spin particles. (If we wanted to exchange even spin meson trajectories, we would project out the odd states.) For the closed string this involves shifting the Regge trajectory so that the lowest particle on it is spin 2.  In both cases, we also need the Regge trajectories to have the physically relevant slopes and intercepts, and require that the incoming and outgoing particles have mass equal to the mass of the proton.

The bosonic open string four-tachyon amplitude\footnote{The external states are not relevant to this analysis; we choose the tachyon amplitude for simplicity.} can be written as a sum of three terms,
\beq
\mathcal{A}^{4}_{o}(s, t, u) = \tilde{\mathcal{A}}_{o}(s, t) + \tilde{\mathcal{A}}_{o}(u, t) + \tilde{\mathcal{A}}_{o}(s, u)~.
\eeq
Each term takes the form
\beq
\tilde{\mathcal{A}}_{o}(x, y) = iC\frac{\Gamma[-a_o(x)]\Gamma[-a_o(y)]}{\Gamma[-a_o(x) - a_o(y)]} \, ,
\eeq

\noindent with $a_o(x) = 1 + \alpha' x$.  $\alpha'$ the Regge slope for bosonic, flat space string theory (so $\alpha'$ is inversely proportional to the string tension).  In order to relate this to vector meson exchange in proton-proton scattering, we will replace the function $a_o(x)$ with the Regge trajectory of vector mesons $\alpha_v(x)$.  This trajectory should satisfy
\beq
J = \alpha_v(m_J^2) = \alpha_{v0} + \alpha_{v}'m_J^2 \, ,
\eeq
and in particular $1 = \alpha_{v0} + \alpha_v'm_v^2$ for the vector meson.  We also need to assume $s + t + u = 4m_p^2$, so that the incoming and outgoing particles are protons.  This implies we should have
\beq
\alpha_v(s) + \alpha_v(t) + \alpha_v(u) = 3 + \alpha_v'(4m_p^2 - 3m_v^2) \equiv \chi_v \, ,
\eeq
which we can use to rewrite dependence on $u$ in terms of $s$ and $t$.  If we then expand these terms around poles corresponding to t-channel exchange, we notice that only $\tilde{\mathcal{A}}(s, t)$ and $\tilde{\mathcal{A}}(u, t)$ possess such poles, and thus $\tilde{\mathcal{A}}(s, u)$ should not really be a part of the Reggeization process.  In addition, in order to only include particles on the vector meson Regge trajectory (those with odd spins), we must use the difference between $\tilde{\mathcal{A}}(s, t)$ and $\tilde{\mathcal{A}}(u, t)$.  In doing so, terms corresponding to the exchanges of even spin particles cancel, while those corresponding to the exchanges of odd spin particles add (see Appendix A for further details).  Thus we write
\beq
\mathcal{A}_{\mathrm{v}}^{4} = iC\frac{\Gamma[\alpha_v(s) + \alpha_v(t) - \chi_v]\Gamma[-\alpha_v(t)]}{\Gamma[\alpha_v(s) - \chi_v]} -  iC\frac{\Gamma[-\alpha_v(s)]\Gamma[-\alpha_v(t)]}{\Gamma[-\alpha_v(s) - \alpha_v(t)]}~.
\eeq
We can then determine the appropriate Reggeization by expanding this expression around the $\alpha_v(t) = 1$ pole (corresponding to exchange of the lightest vector meson) and comparing this with the Regge limit of it.  By doing so, we obtain the prescription
\beq
\frac{1}{t - m_v^2} \hspace{.25in} \rightarrow \hspace{.25in} \alpha_v' \, e^{-\frac{i\pi\alpha_v(t)}{2}} \, \sin\left[\frac{\pi\alpha_v(t)}{2}\right] \, (\alpha_v's)^{\alpha_v(t) - 1} \, \Gamma[-\alpha_v(t)] \, .
\eeq
This, notably, does not depend on the constant $\chi_v$.

If we want to follow the same procedure for the glueball trajectory, we begin instead with the  bosonic closed string four-tachyon amplitude, which can be written as
\beq
\mathcal{A}^4_c(s, t, u) = 2\pi C \, \frac{\Gamma[-a_c(t)]\Gamma[-a_c(s)]\Gamma[-a_c(u)]}{\Gamma[-a_c(s) - a_c(t)]\Gamma[-a_c(s) - a_c(u)]\Gamma[-a_c(t) - a_c(u)]} \, ,
\eeq
where $a_c(x) = 1 + \frac{\alpha' x}{4}$.  We again replace the dependence on the function $a_c(x)$ with a dependence on the physical Regge trajectory of glueballs, $\alpha_g(x)$, which should satisfy
\beq
J = \alpha_g(m_J^2) = \alpha_{g0} + \alpha_{g}'m_J^2 \ .
\eeq
Note that we need to do make the replacement in such a way  that poles exist only for even spin particles, and that the lowest lying particle on the trajectory is a spin-2 glueball with mass $m_g$, with $2 = \alpha_{g0} + \alpha_g'm_g^2$.  This means replacing $a_c(x)$ with $\alpha_g(x) - 2$.  Note that if the incoming and outgoing particles are protons, we will have
\beq
\alpha_g(s) + \alpha_g(t) + \alpha_g(u) = 6 + \alpha_g'(4m_p^2 - 3m_g^2) \equiv \chi_g \, .
\eeq
We can then achieve the desired result by writing
\beq
\mathcal{A}_{\mathrm{g}}^4 = \frac{\Gamma\left[1 - \frac{\alpha_g(t)}{2}\right]\Gamma\left[1 - \frac{\alpha_g(s)}{2}\right]\Gamma\left[1 - \frac{\alpha_g(u)}{2}\right]}{\Gamma\left[2 - \frac{\alpha_g(t)}{2} - \frac{\alpha_g(s)}{2}\right]\Gamma\left[2 - \frac{\alpha_g(t)}{2} - \frac{\alpha_g(u)}{2}\right]\Gamma\left[2 - \frac{\alpha_g(u)}{2} - \frac{\alpha_g(s)}{2}\right]}~.
\eeq

Again we determine the appropriate Reggeization by comparing the pole expansion to the Regge limit, which gives us
\beq
\label{eqn:singlegpres}
\frac{1}{t - m_g^2} \hspace{.25in} \rightarrow \hspace{.25in} \left(\frac{\alpha_g'}{2}\right) \, e^{-\frac{i\pi\alpha_g(t)}{2}} \, \frac{\Gamma\left[3 - \frac{\chi_g}{2}\right]\Gamma\left[1 - \frac{\alpha_g(t)}{2}\right]}{\Gamma\left[2 - \frac{\chi_g}{2} + \frac{\alpha_g(t)}{2}\right]} \, \left(\frac{\alpha_g' s}{2}\right)^{\alpha_g(t) - 2} \, ,
\eeq
where the dependence on $\chi_g$ is introduced when we replace dependence on $u$ with dependence on $s$ and $t$. Note that in this case, unlike for the open string amplitude, $\chi_g$ does not end up canceling out.

\subsection{Review of the 5-Closed-String Reggeization Process}

Having determined a Reggeization procedure for meson and glueball propagators appropriate to elastic proton-proton scattering, we now turn to the more complicated case of central production, where there is a (possibly Reggeized) central vertex.  We begin with the Reggeization process for the glueball propagators in central $\eta$ production, analyzed previously in \cite{ADHM}. Our starting point is the dual $2\rightarrow 3$ process: the closed bosonic string 5-tachyon amplitude as analyzed in \cite{Herzog}, which can be written as an integral over two copies of the complex plane:
\beq
\mathcal{A}^{5}_c = C\int d^2ud^2v \, |u|^{-2a_c(t_1) - 2}|v|^{-2a_c(t_2) - 2}|1-u|^{-2a_c(s_1) - 2}|1-v|^{-2a(s_2) -2}|1 - uv|^{2a_c(s_1) + 2a_c(s_2) - 2a_c(s) - 2} \, .
\eeq
This amplitude is difficult to compute in closed form, though it can be written in terms of generalized hypergeometric functions \cite{BBMRR, BFM}.  However, we will examine it instead by taking the Regge limit under two different conditions, depending on the value of $\frac{\alpha'\mu}{4}$, where $\mu = \frac{s_1s_2}{s}$ is the kinematic parameter held constant in the Regge limit and $\alpha'$ is again the Regge slope for bosonic, flat space string theory.  If $\frac{\alpha'\mu}{4}$ is large, then we obtain
\beq
\mathcal{A}^{5}_c \approx 4\pi^2 C \, \left(\frac{-i\alpha' s_1}{4}\right)^{2a_c(t_1)}\left(-\frac{i\alpha' s_2}{4}\right)^{2a_c(t_2)} \, \frac{\Gamma[-a_c(t_1)]\Gamma[-a_c(t_2)]}{\Gamma[a_c(t_1) + 1]\Gamma[a_c(t_2) + 1]} \, ,
\eeq
which we recognize as essentially the product of two 4-closed-string amplitudes in the Regge limit.  On the other hand, if $\frac{\alpha'\mu}{4}$ is small then we obtain
\beq
\mathcal{A}^{5}_c \approx -4\pi^2 C \, \Bigg\{\left(\frac{s}{s_2}\right)^{2a_c(t_1)}\left(\frac{-i\alpha' s_2}{4}\right)^{2a_c(t_2)}\frac{\Gamma[-a_c(t_1)]\Gamma[a_c(t_1) - a_c(t_2)]}{\Gamma[1 + a_c(t_1)]\Gamma[1 + a_c(t_2) - a_c(t_1)]}
\eeq
$$
+ \left(\frac{s}{s_1}\right)^{2a_c(t_2)}\left(\frac{-i\alpha' s_1}{4}\right)^{2a_c(t_1)}\frac{\Gamma[-a_c(t_2)]\Gamma[a_c(t_2) - a_c(t_1)]}{\Gamma[1 + a_c(t_2)]\Gamma[1 + a_c(t_1) - a_c(t_2)]}\Bigg\} \, .
$$
When we modify this to take into account a physical Regge trajectory of glueballs, we will be concerned with the product $\frac{\alpha_g'\mu}{2}$.  The value of $\mu$ is primarily determined by the mass of the centrally produced eta meson: $\mu \sim m_{\eta}^2$, and we know $\alpha_g' = 0.3 \ \mathrm{GeV}^{-2}$ based on fitting to proton-proton scattering data \cite{DHM}, which gives $\frac{\alpha_g'\mu}{2} \sim 0.04$.  This is clearly more consistent with using the second expression.  If we were following the same procedure as we did for the 4-string amplitudes, we would now want to modify this expression to fit the physical Regge trajectory and then compare it to a pole expansion.  The fact that we do not have the amplitude calculated exactly in closed form makes this difficult.  However, we can instead assume that in the large $\frac{\alpha_g'\mu}{2}$ limit we simply have two copies of the rule given  \eqref{eqn:singlegpres}, and  use this to deduce the rule for small $\frac{\alpha_g'\mu}{2}$.  This leads to the prescription
\beq
\frac{1}{(t_1 - m_g^2)(t_2 - m_g^2)} \hspace{.25in} \rightarrow \hspace{.25in} -\frac{1}{s^2}\mathcal{P}_g = -\frac{1}{s^2}e^{-\frac{i\pi[\alpha_g(t_1) + \alpha_g(t_2)]}{4}}\Gamma\left[3 - \frac{\chi_g}{2}\right]^2 \left[\frac{\alpha_g' s}{2}\right]^{\frac{\alpha_g(t_1) + \alpha_g(t_2)}{2}}
\eeq
$$
\times \left\{\left[\frac{\alpha_g'\mu}{2}\right]^{\frac{\alpha_g'(t_2 - t_1)}{2}} \, \frac{e^{\frac{i\pi \alpha_g'(t_1 - t_2)}{4}}\Gamma\left[1 - \frac{\alpha_g(t_1)}{2}\right]\Gamma\left[\frac{\alpha_g'(t_1 - t_2)}{2}\right]}{\Gamma\left[\frac{\alpha_g(t_1)}{2} + 2 - \frac{\chi_g}{2}\right]\Gamma\left[\frac{\alpha_g'(t_2 - t_1)}{2} + 3 - \frac{\chi_g}{2}\right]}  + \right.
$$
$$
\left. \left[\frac{\alpha_g'\mu}{2}\right]^{\frac{\alpha_g'(t_1 - t_2)}{2}} \, \frac{e^{\frac{i\pi \alpha_g'(t_2 - t_1)}{4}}\Gamma\left[1 - \frac{\alpha_g(t_2)}{2}\right]\Gamma\left[\frac{\alpha_g'(t_2 - t_1)}{2}\right]}{\Gamma\left[\frac{\alpha_g(t_2)}{2} + 2 - \frac{\chi_g}{2}\right]\Gamma\left[\frac{\alpha_g'(t_1 - t_2)}{2}  + 3 - \frac{\chi_g}{2}\right]}\right\} \, .
$$
Note that there is some ambiguity here. In proton-proton scattering we introduced the dependence on the factor $\chi_g$ to account for the fact that the incoming and outgoing particles had to have the masses of protons; it arose from the mass-shell condition on the Mandelstam variables.  Here we have assumed that some of the gamma function terms must be shifted in a similar way.  However, there is no real reason to suppose that exactly the same factor of $\chi_g$ should be introduced: in fact we would naively expect that it should now also depend on the mass of the centrally produced $\eta$ meson.  


It is also worth mentioning that the procedure used here (and for the 5-open-string amplitude) is different than that used for the 4-string amplitudes.  In those cases we were able to start from string amplitudes written in closed form, in terms of gamma functions.  Here, we chose to apply the Regge limit while the amplitude was still in integral form.  There is some evidence that this distinction might matter: applying a similar integral-based technique to the 4-closed-string amplitude generates a result that differs in the value of $\chi_g$ \cite{un}.  This is another reason to consider the choice of $\chi_g$ in the above expression to be slightly suspect.  However, changing the value of $\chi_g$ should not change the result significantly, provided it is negative and $\mathcal{O}(1)$, so we will use the same value as before.

\subsection{The 5-Open-String Reggeization Process}

We turn finally to our new result: the procedure for Reggeizing the vector meson propagators connected to a central vertex.  This time we begin with the open bosonic string 5-tachyon amplitude, which can be written as
\beq
\mathcal{A}_o^5 = 2iC\int_{-\infty}^{\infty} \int_{-\infty}^{\infty} dx \, dy |x|^{-1 - a_o(t_1)}|y|^{-1 - a_o(t_2)}|1 - x|^{-1 - a_o(s_1)}|1 - y|^{-1 - a_o(s_2)}|1 - xy|^{a_o(s_1) + a_o(s_2) - a_o(s)} \, .
\eeq
In the Regge limit, the integral is dominated by regions near four points, given by $x \sim \pm \frac{1}{s_1}$ and $y \sim \pm \frac{1}{s_2}$.  We can therefore split up the region of integration into four separate pieces, each of which contains one of these points.  These four terms are analogous to the two expressions $\tilde{A}_o(s, t)$ and $\tilde{A}_o(u, t)$ that contributed to the Reggeization process for elastic proton-proton scattering.  Just as was required there, we will need to use a particular linear combination of these terms that allows us to project out poles associated with the exchange of even-spin particles.  These integrals, like the ones we encountered in the 5-closed-string amplitude, are difficult to compute in closed form, so we again expand in the large $\alpha'\mu$ and small $\alpha'\mu$ limits.  Following this procedure for the large $\alpha'\mu$ limit gives us
\beq
\mathcal{A}_{o, \mathrm{odd}}^{5} \approx 8iC \, e^{-\frac{i\pi a_o(t_1)}{2}} \, e^{-\frac{i\pi a_o(t_2)}{2}} \, \sin\left[\frac{\pi a_o(t_1)}{2}\right] \, \sin\left[\frac{\pi a_o(t_2)}{2}\right] \, (\alpha' s_1)^{a_o(t_1)} \, (\alpha' s_2)^{a_o(t_2)} \, \Gamma[-a_o(t_1)]\Gamma[-a_o(t_2)]
\eeq
which again is straightforward to identify as being essentially two copies of what we had for the 4-string amplitude.  On the other hand,  small $\alpha'\mu$ gives
\beq
\mathcal{A}^5_{o, \, \mathrm{odd}} \approx -8iC \, e^{-\frac{i\pi(a_o(t_1) + a_o(t_2)}{4}} \, \cos\left[\frac{\pi\alpha'(t_2 - t_1)}{2}\right] \, (\alpha' s)^{\frac{a_o(t_1) + a_o(t_2)}{2}}
\eeq
$$
\times \left\{e^{\frac{i\pi\alpha'(t_1 - t_2)}{4}} \, (\alpha'\mu)^{\frac{\alpha'(t_2 - t_1)}{2}} \, \sin\left[\frac{\pi a_o(t_1)}{2}\right] \, \Gamma[-a_o(t_1)] \, \Gamma[\alpha'(t_2 - t_1)] \right.
$$
$$
+ \left. e^{\frac{i\pi\alpha'(t_2 - t_1)}{4}} \, (\alpha'\mu)^{\frac{\alpha'(t_1 - t_2)}{2}} \, \sin\left[\frac{\pi a_o(t_2)}{2}\right] \, \Gamma[-a_o(t_2)] \, \Gamma[\alpha'(t_1 - t_2)] \right\} \, .
$$

When we modify this result to take into account the physical Regge trajectories, we will be concerned with the product $\alpha'_v\mu$.  We know that $\alpha_v' = 0.89 \ \mathrm{GeV}^{-2}$ from a linear fit to known meson masses \cite{PDG}, which gives $\alpha_v'\mu \sim 0.27$.  This is certainly more consistent with the ``small $\mu$'' limit than with the ``large $\mu$'' limit, though the approximation this leads to here will not be as good as the one used for the closed string case.  The ``small $\mu$'' limit comes from expanding and keeping the leading term in a hypergeometric function, and if we use this treatment we expect the next term to constitute a roughly $12\%$ correction.

Following the same basic procedure we used for the 5-closed-string amplitude,  we obtain the rule
\beq
\frac{1}{(t_1 - m_v^2)(t_2 - m_v^2)} \hspace{.25in} \rightarrow \hspace{.25in} i\frac{\alpha_v'}{s}\mathcal{P}_v = i\frac{\alpha_v'}{s} \, e^{-\frac{i\pi(\alpha_v(t_1) + \alpha_v(t_2)}{4}} \, \cos\left[\frac{\pi\alpha'_v(t_2 - t_1)}{2}\right] \, (\alpha_v' s)^{\frac{\alpha_v(t_1) + \alpha_v(t_2)}{2}}
\eeq
$$
\times \left\{e^{\frac{i\pi\alpha_v'(t_1 - t_2)}{4}} \, (\alpha_v'\mu)^{\frac{\alpha_v'(t_2 - t_1)}{2} - 1} \, \sin\left[\frac{\pi \alpha_v(t_1)}{2}\right] \, \Gamma[-\alpha_v(t_1)] \, \Gamma[\alpha'_v(t_1 - t_2)] \right.
$$
$$
+ \left. e^{\frac{i\pi\alpha_v'(t_2 - t_1)}{4}} \, (\alpha_v'\mu)^{\frac{\alpha_v'(t_1 - t_2)}{2} - 1} \, \sin\left[\frac{\pi \alpha_v(t_2)}{2}\right] \, \Gamma[-\alpha_v(t_2)] \, \Gamma[\alpha_v'(t_2 - t_1)] \right\} \, .
$$
Recall that when we modified the 4-closed-string amplitude for proton scattering, we found that our Reggeization procedure depended on a parameter $\chi_g$ that involved the masses of the proton and the spin-2 glueball, and appeared as a consequence of  the mass-shell condition.  When we modified the 5-closed-string amplitude, we  assumed that a similar parameter must appear, though we had less direct information as to what exactly it should be.  In comparison, the modification of the 4-open-string amplitude did not end up depending on the analogous factor $\chi_v$.  Consequently, we have not introduced any such factor here, either.

Our choice is again slightly ambiguous, given that the mass-shell condition on the Mandelstam variables will be somewhat different for the central production process than for  elastic scattering.  However, since no modification involving this fact was required for the elastic proton-proton scattering case, it seems reasonable to assume that none is required for central production, either.  In fact the above result may be somewhat more reliable than that obtained for the exchange of glueballs, since it does not involve introducing a parameter whose exact value we do not know.  Furthermore, in the end we will find that the contributions from the exchanges of the $\rho$ and $\omega$ Regge trajectories are larger than that for the glueballs at the energies we are interested in.

Of particular interest in the above expression is the dependence on $\mu$.  We expect the differential cross section to be maximized for small values of $t_1 - t_2$, which means the amplitude scales roughly with $(\alpha_v'\mu)^{-1}$.  This is a substantial dependence on $\mu$ which does not arise in the Reggeization of the glueballs.  Most importantly, it introduces significant additional dependence on $\theta_{34}$ into the result, which as will be discussed in the next section, is not consistent with experimental results.

\section{\label{Results} Predictions for $\eta$ Central Production}

\begin{table}
\begin{center}
\begin{tabular}{|c|c|c|}
\hline
\hline
 {\bf parameter} & {\bf value} & {\bf source} \\
\hline
\hline
$m_\eta$ & 0.548 GeV & known experimental value \cite{PDG} \\
\hline
$m_p$ & 0.938 GeV& known experimental value \cite{PDG} \\
\hline
$\alpha_{v0}$ & 0.456 & value based on linear fit to experimentally known masses \cite{PDG} \\
\hline
$\alpha_{v}'$ & 0.886 GeV$^{-2}$ & value based on linear fit to experimentally known masses \cite{PDG} \\
\hline
$\alpha_{g0}$ & 1.08 & value based on linear fit to proton-proton scattering data \cite{DHM} \\
\hline
$\alpha_g'$ & 0.290 GeV$^{-2}$ & value based on fit to proton-proton scattering data \cite{DHM} \\
\hline
$M_{dg}$ & 1.17 GeV & the proton as a 4-dimensional skyrmion in the Sakai-Sugimoto model \cite{DHM} \\
\hline
$M_{dv}$ & 0.843 GeV & empirically determined dipole mass \cite{L.A.}  \\
\hline
$g_{\rho\rho\eta}$ & -1.9 GeV$^{-1}$ & the Sakai-Sugimoto dual model in \cite{Domokos:2009cq} \\
\hline
$g_{\omega\omega\eta}$ & -1.9 GeV$^{-1}$ & the Sakai-Sugimoto dual model in \cite{Domokos:2009cq} \\
\hline
$\hat{g}_{hh\eta}$ & 0.0222 GeV$^{-1}$ & Sakai-Sugimoto calculation in \cite{ADHM} \\
\hline
$\tilde{g}_{hh\eta}$ & 0.0482 GeV$^{-3}$ & Sakai-Sugimoto calculation in \cite{ADHM} \\
\hline
$\lambda_{pp\rho}$ & -3.59 & Sakai-Sugimoto effective fermion calculation in \cite{Hong:2007ay} \\
\hline
$\tilde{\lambda}_{pp\rho}$ & 4.0 GeV$^{-1}$ & Sakai-Sugimoto effective fermion calculation in section (\ref{lambda}) \\
\hline
$\lambda_{pp\omega}$ & -12.53 & Sakai-Sugimoto effective fermion calculation in \cite{Hong:2007ay} \\
\hline
$\check{\lambda}_{pph}$ & 9.02 GeV$^{-1}$ & effective fermion fields in the Sakai-Sugimoto model, in \cite{Domokos:2010ma} \\
\hline\hline
\end{tabular}
\caption{\label{table} Parameters appearing in the differential cross section.}
\end{center}
\end{table}

Combining the results of the previous sections, we can now compute the total cross section for $\eta$ production, and perform a Monte Carlo simulation to estimate  the differential cross section.  We can then discuss these results in comparison with experimental data.  Our primary focus will be studying the $pp\rightarrow pp\eta$ process at $\sqrt{s} = 29.1 \ \mathrm{GeV}$, so as to compare the results with data from the WA102 experiment run at that energy \cite{WA102, Kirk}.  We are interested in the total cross section at that energy, as well as the differential cross section as a function of angle $\theta_{34}$, and of $t_1$ and $t_2$.  We will also make predictions for a range of $s$ values, where we can examine the competing effects of Pomeron and Reggeon exchange.

Combining the results of sections \ref{Feynman} and \ref{Reggeization}, the Reggeized differential cross section is

\beq
\frac{d^3\sigma}{dt_1 \, dt_2 \, d\theta_{34}} = \frac{t_1t_2\sin^2\theta_{34}}{(4\pi)^4 \, s^2} \, \ln \left(\frac{s}{\mu}\right) \ \Bigg\{ 16\alpha_v'^2\tilde{\lambda}^2_{pp\rho}g_{\rho\rho\eta}^2(4\tilde{\lambda}^2_{pp\rho}t_1t_2 - \lambda^2_{pp\rho}t_1 - \lambda^2_{pp\rho}t_2)A_v(t_1)A_v(t_2)\Big|\mathcal{P}_v\Big|^2~.
\eeq
$$
+ \left|\check{\lambda}_{pph}^2A_g(t_1)A_g(t_2)\Big(\hat{g}_{hh\eta} - \tilde{g}_{hh\eta}\sqrt{t_1t_2}\cos\theta_{34}\Big)\mathcal{P}_g
+ 2i\alpha_v'(\lambda_{pp\rho}^2g_{\rho\rho\eta} + \lambda_{pp\omega}^2g_{\omega\omega\eta})A_v(t_1)A_v(t_2)\mathcal{P}_v\right|^2\Bigg\}
$$

The values of all parameters appearing here are shown in table \ref{table}.  All of the coupling constants are derived from low energy calculations in the Sakai-Sugimoto model.  The proton and $\eta$ particle masses are taken from experimental data \cite{PDG}, and the slope and intercept of the vector meson Regge trajectory are derived from a linear fit to known experimental mass values.  The slope and intercept of the Pomeron trajectory are derived from a fit to experimental proton-proton scattering data.  In evaluating the total and differential cross sections, we will restrict the $t$-values to the range $-0.6 \ \mathrm{GeV}^2 < t_1, t_2 < 0$, because outside of this range we expect significant perturbative QCD effects.

The total cross section computed using this model at $\sqrt{s} = 29.1 \ \mathrm{GeV}$ is
\beq
\sigma_{\mathrm{tot}} = 236 \ \mathrm{nb}~,
\eeq
which is roughly an order of magnitude smaller than the experimental value
\beq
\sigma_{\mathrm{tot}, \ \mathrm{exp}} = 3859 \ \mathrm{nb}~,
\eeq
However, it should be remembered that coupling constants and masses computed in the Sakai-Sugimoto model are generally only accurate to around 15\%, and our ``small $\mu$'' approximation in the Reggeization procedure should also introduce an error of about 12\%. There is also evidence based on recent experimental work involving photoproduction of the $f_1(1285)$ meson that the Sakai-Sugimoto model, with the values of $m_{\rho}$ and $f_{\pi}$ used to fix free parameters, may systematically underestimate coupling constants \cite{CLAS}.  If all of the values of coupling constants were too small by roughly $10\%-15\%$, this could lead to the total cross section being suppressed by a factor of $1/3$ or more.

\begin{figure}
\begin{center}
\includegraphics{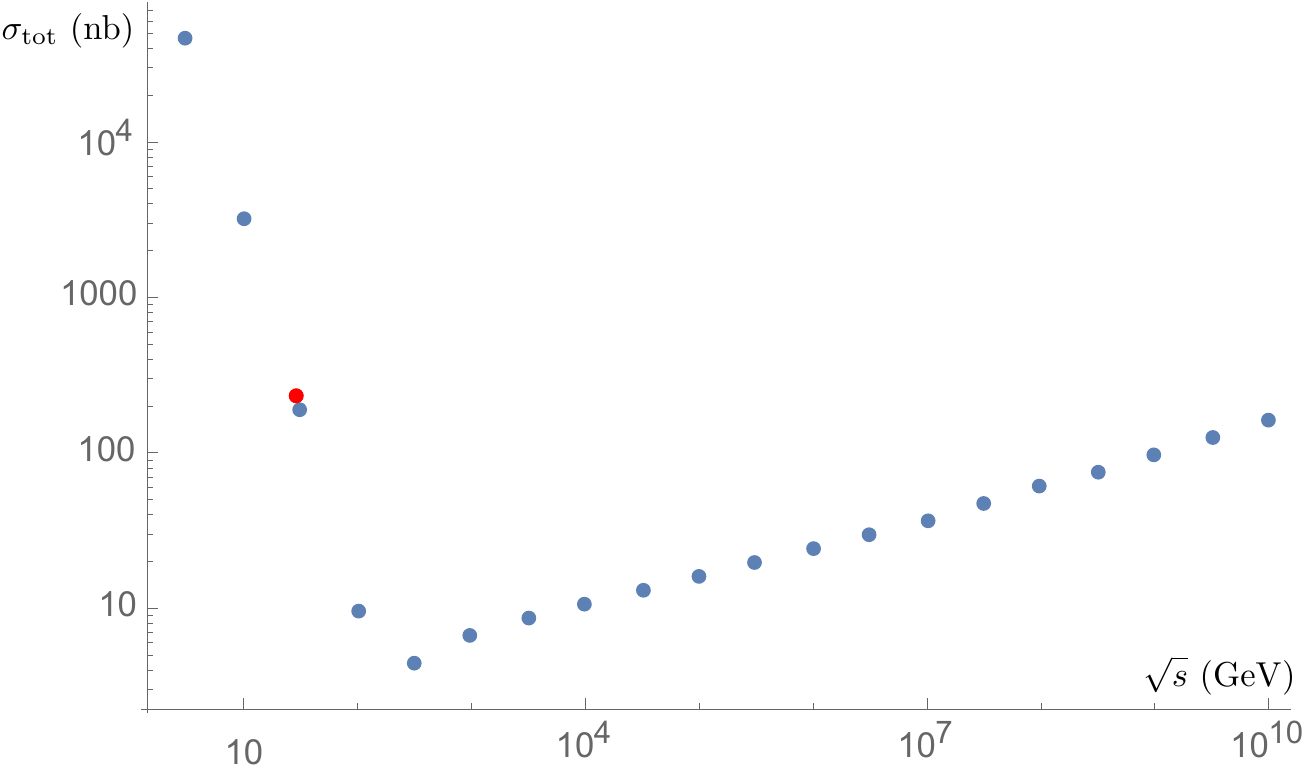}
\caption{\label{total} The total cross section as a function of $\sqrt{s}$.  The red dot indicates the value $\sqrt{s} = 29.1 \ \mathrm{GeV}$.}
\end{center}
\end{figure}

We can also consider a range of different center-of-mass energies, $\sqrt{s}$, shown in figure \ref{total}.  As expected, at smaller values of $\sqrt{s}$, vector meson exchange dominates; at larger values of $\sqrt{s}$, glueball exchange dominates.  For mid-range values, these two processes tend to cancel against each other, leading to a smaller total cross section.  The growth of the total cross section with increasing $\sqrt{s}$ once the glueball dominates is slow, which makes sense given that the dominate scaling behavior is $\sigma \sim s^{2(\alpha_{g0} - 1)}$, which implies $\sigma \sim (\sqrt{s})^{0.32}$.  The value we are using to compare to data is shown in red: it is clear from its location that in this regime the Reggeon exchange is the dominant effect.

It is worth remembering that we chose to fix the Regge trajectory parameters for the Pomeron by using a fit to experimental high energy elastic proton-proton scattering data.  This is not completely consistent with the value of the glueball mass $m_g$ that is found by using the Sakai-Sugimoto model, and if we had used it to help fix the value of $\alpha_{g0}$, we would have gotten a larger intercept value.  This in turn would imply a larger Pomeron contribution at the energy $\sqrt{s} = 29.1 \ \mathrm{GeV}$, along with a shift in the energy at which Pomeron exchange begins to play a significant role.  As long as we stick with the regime where Reggeons dominate, however, this affects the results by only a small amount.


\begin{figure}
\begin{center}
\includegraphics{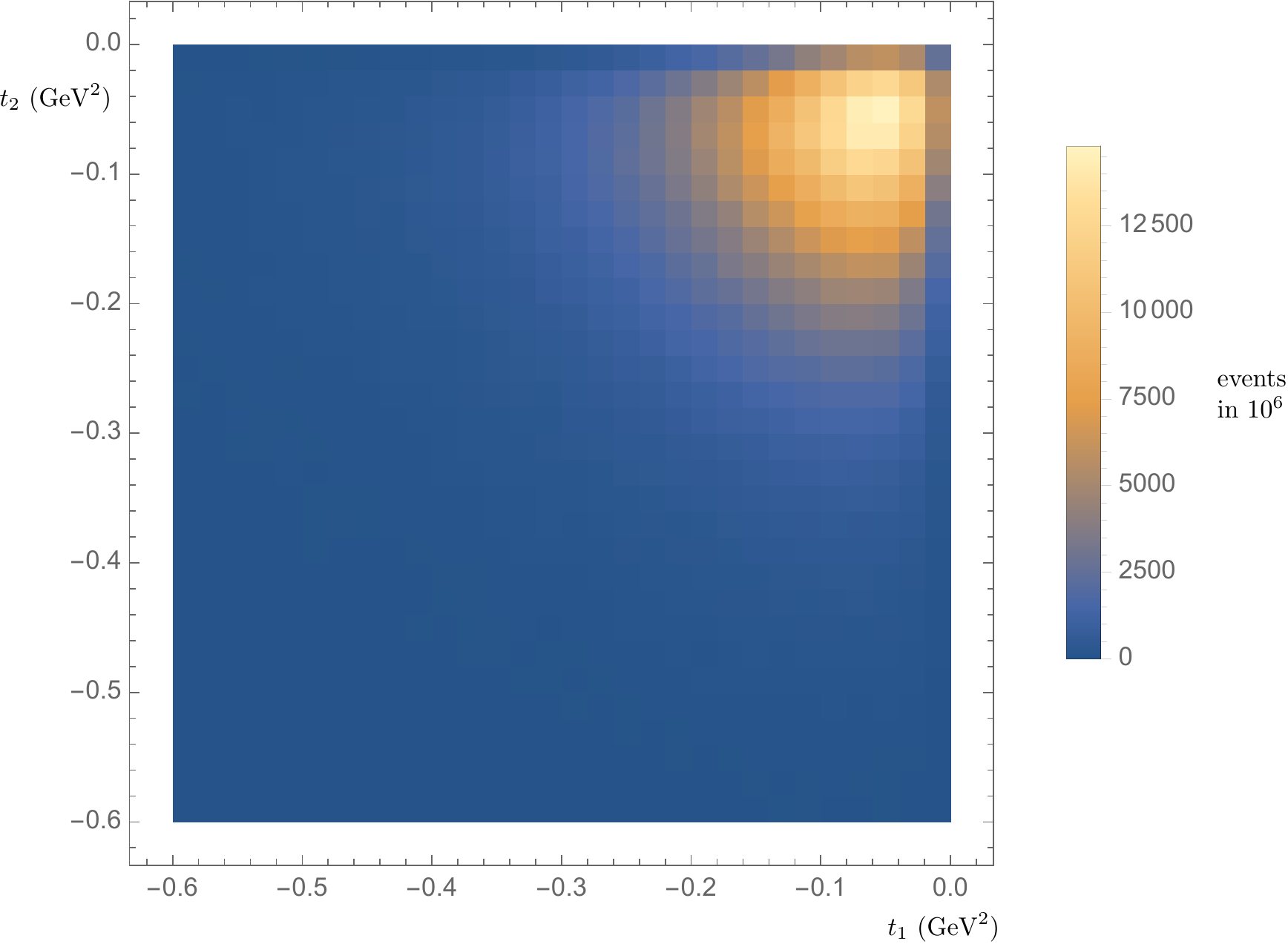}
\caption{\label{thist} A histogram showing the number of events for different values of $t_1$ and $t_2$, with $10^6$ total events.}
\end{center}
\end{figure}

In figure \ref{thist} we show the distribution of simulated events over different values of $t_1$ and $t_2$.  Here we find that the differential cross section is maximized around $t_1, t_2 \approx -0.05 \ \mathrm{GeV}^2$.  The number of events drops off significantly for larger values of $|t_1|$ and $|t_2|$, which supports our restriction to the range $-0.6 \ \mathrm{GeV}^2 < t_1, t_2 < 0$.

\begin{figure}
\begin{center}
\includegraphics{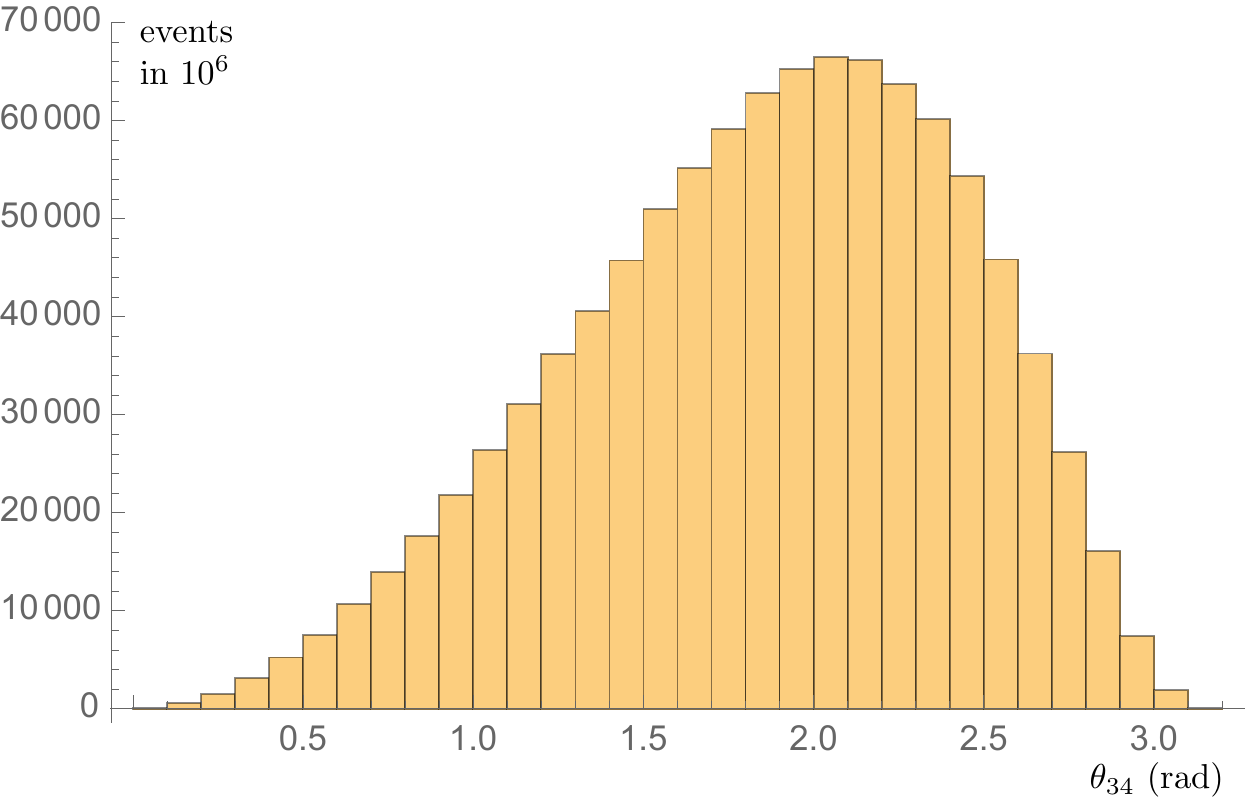}
\caption{\label{thetahist} A histogram showing the number of events as a function of $\theta_{34}$, with $10^6$ total events.}
\end{center}
\end{figure}

Figure \ref{thetahist} shows the dependence of the differential cross section at $\sqrt{s}=29.1$ GeV as a function of $\theta_{34}$.  This differential cross section has a maximum around $\theta_{34} = 1.9$ radians.  Here we see possible disagreement with the experimental result, which gives a result roughly symmetric around $\theta_{34} = \frac{\pi}{2}$, consistent with an amplitude proportional to $\sin^2\theta_{34}$.  The additional angular dependence in our model arises from two places: the structure of the coupling constants for glueball exchange, and the dependence of the cross section of $\mu$, which is particularly strong for Reggeon exchange. At  $\sqrt{s} = 29.1 \ \mathrm{GeV}$ the dominant contribution is  Reggeon exchange, so the $\theta_{34}$ behavior of glueball exchange is not the primary effect here. This implies that the source of our disagreement may lie in the Reggeization procedure.

\section{\label{Conclusion} Conclusions and Future Directions}

In this work, we built a holographic model for the central production of $\eta$ mesons in proton-proton scattering via the exchange of Reggeons and Pomerons.  Our model was constructed in two stages: (1) the calculation of a low-energy cross section for central $\eta$ production via the exchange of mesons and glueballs, and (2) the generalization of this cross section to include the effects of exchanging full Regge trajectories (as opposed to single particles).  Given this Regge regime prediction, we compared our results to experimental data and found that the total cross section was almost an order of magnitude smaller than the measured value.

Our differential cross section showed dependence on the Mandelstam variables $t_1$ and $t_2$ consistent with our expectations: the magnitude of the cross section fell off sharply for larger values of $|t_1|$ and $|t_2|$, with comparatively insignificant contributions outside the range $-0.6 \ \GeV < t_1, t_2 < 0$, where our model best applies.  More interesting was the dependence showed on $\theta_{34}$.  The experimental data is clearly consistent with the dominant effect being the factor of $\sin^2\theta_{34}$ associated with natural parity violation, but does not show strong evidence for any other features.  In contrast, our model shows a significant asymmetry in its dependence on $\theta_{34}$ that can be traced back to our Reggeization procedure for the vector mesons.

This suggests that our Reggeization procedure may not be ideal.  In fact, our method involved taking the Regge limit of the string amplitudes while they were still in integral form.  An alternate approach would be to write the 5-tachyon amplitude in terms of hypergeometric functions in a form with manifest symmetry under exchanges of the external particles, and only then take the Regge limit.  This is more consistent with what has been done for elastic proton-proton scattering previously, and it might produce a somewhat different result, perhaps more consistent with the experimental data.

However, some of the discrepancies could arise in part from model-dependent uncertainties, and do not necessarily indicate a flaw in the Reggeization procedure (or gauge-string duality).  In choosing our Regge trajectory parameters, we used a linear fit to known meson masses for the Reggeon and a fit to very high energy elastic proton-proton scattering for the Pomeron.  The slope and intercept for the Pomeron trajectory obtained this way are not in close agreement with the Sakai-Sugimoto model calculation for the mass of the lowest spin-2 glueball state: the trajectory would give this a value of $m_g = 1.781$ GeV, while the Sakai-Sugimoto model gives a value of $m_g = 1.485$ GeV.  We could instead have used the Sakai-Sugimoto glueball mass along with a fit value for the slope of the trajectory to estimate the intercept; this would have led to a larger intercept, and thus to a larger contribution to the total process by the glueballs.

In order to compute the low energy cross section via gauge-string duality, furthermore, we used the well-known Sakai-Sugimoto construction in the supergravity limit.  The couplings computed in AdS/QCD models are only accurate to 10\%-15\%, and the terms in our cross section each carry six factors of coupling constants, all told, making our results reasonably consistent with experimental values.  Interestingly,  our results also corroborate the recent suggestion from experimental measurement of $f_1$ photoproduction \cite{CLAS}, that the Sakai-Sugimoto model systematically underestimates the values of low-energy coupling constants involved in these Regge regime processes. It would be interesting to explore corrections to these parameters as one departs from the strict supergravity limit of vanishing string length.

Over all, our model for central $\eta$ production is in weak agreement with the experimental data, but it may still be consistent with the quality of agreement between the Sakai-Sugimoto model and experimental results that have been obtained elsewhere.  Discrepancies between the two may add to our understanding of the limitations of AdS/QCD models, and perhaps to ways in which they might be improved.

\begin{acknowledgements}
SKD was supported by NSF grant PHY-1214302, and thanks NYU's Center for Cosmology and Particle Physics, where much of her work on this manuscript was done. NM thanks Richard Brower for useful discussions.
\end{acknowledgements}

\begin{appendix}

\section{Reggeization Procedures \label{regge}}

\subsection{Open 4-String Process}

\subsubsection{In Bosonic String Theory}

If we write down the open string 4-tachyon amplitude, this is
\beq
\mathcal{A}^{4}_{o}(s, t, u) = \tilde{\mathcal{A}}_{o}(s, t) + \tilde{\mathcal{A}}_{o}(u, t) + \tilde{\mathcal{A}}_{o}(s, u) \, ,
\eeq
where each term is in the form
\beq
\tilde{\mathcal{A}}_{o}(x, y) = iC\frac{\Gamma[-a_o(x)]\Gamma[-a_o(y)]}{\Gamma[-a_o(x) - a_o(y)]} \, .
\eeq
Here, $a_o(x) = 1 + \alpha' x$ is the leading Regge trajectory of open string states, so that the mass of the open string tachyon is $m_{To}^2 = -\frac{1}{\alpha'}$.  This means that the mass shell condition guarantees $s + t + u = 4m_{To}^2 = -\frac{4}{\alpha'}$ and therefore
\beq
a_o(s) + a_o(t) + a_o(u) = -1 \, .
\eeq

Suppose we begin with a t-channel propagator associated to the exchange of a single particle (string state), and we want to replace this with a propagator that takes into account the exchange of a whole Regge trajectory, in the Regge limit.  What we want to do is compare the expansion of the string amplitude around the appropriate pole with the same amplitude's Regge limit.  We find that of the three terms above, only $\tilde{A}_o(s, t)$ and $\tilde{\mathcal{A}}_o(u, t)$ have the right behavior in the Regge limit, and only these have poles in the $t$-channel.  Thus we can ignore $\tilde{\mathcal{A}}_{o}(s, u)$.  The other two each have poles associated with $a_0(t) = n$, where $n$ is any non-negative integer.  The pole expansion then gives
\beq
\tilde{\mathcal{A}}_o(s, t)\Bigg|_{a_o(t) = n} \ \approx \ iC \, \frac{(a_o(s) + 1)\cdots(a_o(s)+n)}{n!(n - a_o(t))}, \hspace{.5in} \tilde{\mathcal{A}}_o(u, t)\Bigg|_{a_o(t) = n} \ \approx \ (-1)^{n} \times iC \, \frac{(a_o(s) + 1)\cdots(a_o(s)+n)}{n!(n - a_o(t))}
\eeq
while the Regge limits are
\beq
\tilde{\mathcal{A}}_o(s, t)\Bigg|_{s\gg t} \ \approx \ iC \, e^{-i\pi a_o(t)} \, \left(\alpha' s\right)^{a_o(t)} \, \Gamma[-a_o(t)],
\hspace{.5in}
\tilde{\mathcal{A}}_o(u, t)\Bigg|_{s\gg t} \ \approx \ iC \, \left(\alpha' s\right)^{a_o(t)} \, \Gamma[-a_o(t)] \, .
\eeq

Suppose we want to model the exchange of a Reggeon consisting just of the odd spin string states, using the vector (spin 1) particle propagator.  Then it's clear we must take the difference between $\tilde{\mathcal{A}}_o(s, t)$ and $\tilde{\mathcal{A}}_o(u, t)$ as our starting point: this projects out the even spin particles.  Thus we could write
\beq
\mathcal{A}^{4}_{o, \, \mathrm{odd}} = \mathcal{A}_o(u, t) - \mathcal{A}_o(s, t)
\eeq
and comparing the expansion around $a_o(t) = 1$ to the Regge limit gives
\beq
\frac{1}{t - m_1^2} \hspace{.25in} \rightarrow \hspace{.25in} -\alpha' \, e^{-\frac{i\pi a_o(t)}{2}} \, \sin\left[\frac{\pi a_o(t)}{2}\right] \, (\alpha' s)^{a_o(t) - 1} \, \Gamma[-a_o(t)] \, .
\eeq

\subsubsection{Modifications for Physical Reggeons and Proton-Proton Scattering}

If instead we want to use physical Reggeons, which will consist of a trajectory of odd spin particles for which the lowest spin particle is a vector meson, we need to start with a physical trajectory
\beq
\alpha_v(x) = \alpha_{v0} + \alpha_v'x
\eeq
such that the meson trajectory (either $\rho$ or $\omega$) has
\beq
J = \alpha_{v0} + \alpha_v'm_J^2 = \alpha_v(m_J^2) \, ,
\eeq
If we are modeling elastic proton-proton scattering we also need to assume $s + t + u =4m_p^2$ and therefore
\beq
\alpha_v(s) + \alpha_v(t) + \alpha_v(u) =  3 + \alpha_v'(4m_p^2 - 3m_v^2) \equiv \chi_v
\eeq
We then should start with a ``string inspired'' amplitude, which will have poles where $\alpha_v(t)$ is an odd integer, the lowest of which will correspond to the vector meson.  Written in terms of just $s$ and $t$, we will say this is
\beq
\mathcal{A}^{4}_v =iC\frac{\Gamma[\alpha_v(s) + \alpha_v(t) - \chi_v]\Gamma[-\alpha_v(t)]}{\Gamma[\alpha_v(s) - \chi_v]} -  iC\frac{\Gamma[-\alpha_v(s)]\Gamma[-\alpha_v(t)]}{\Gamma[-\alpha_v(s) - \alpha_v(t)]} \, .
\eeq
A pole expansion of this expression around the $\alpha_v(t) = 1$ pole then gives
\beq
\mathcal{A}^{4}_v\Bigg|_{\alpha_v(t) = 1} \approx \frac{iC(2\alpha_v(s) + 1 - \chi_v)}{1 - \alpha_v(t)}
\eeq
and the Regge limit is
\beq
\mathcal{A}^{4}_v\Bigg|_{s \gg t} \approx iC\left(1 - e^{-i\pi\alpha_v(t)}\right) \, \left(\alpha_v's\right)^{\alpha_v(t)} \, \Gamma[-\alpha_v(t)] \, .
\eeq
Putting these pieces together leads to the Reggeization prescription
\beq
\frac{1}{t - m_v^2} \hspace{.25in} \rightarrow \hspace{.25in} -\alpha_v' \, e^{-\frac{i\pi\alpha_v(t)}{2}} \, \sin\left[\frac{\pi\alpha_v(t)}{2}\right] \, (\alpha_v' s)^{\alpha_v(t) - 1} \, \Gamma[-\alpha_v(t)] \, .
\eeq

\subsection{Closed 4-String Process}

\subsubsection{In Bosonic String Theory}

Now we want to repeat this process for closed Bosonic strings, where the 4-tachyon amplitude is
\beq
\mathcal{A}^4_c(s, t, u) = 2\pi C \, \frac{\Gamma\left[-\frac{a_c(t)}{2}\right]\Gamma\left[-\frac{a_c(s)}{2}\right]\Gamma\left[-\frac{a_c(u)}{2}\right]}{\Gamma\left[-\frac{a_c(s)}{2} - \frac{a_c(t)}{2}\right]\Gamma\left[-\frac{a_c(s)}{2} - \frac{a_c(u)}{2}\right]\Gamma\left[-\frac{a_c(t)}{2} - \frac{a_c(u)}{2}\right]} \, ,
\eeq
Here, $a_c(x) = 2 + \frac{\alpha' x}{2}$ is the leading Regge trajectory of closed string states, which should only include even spin particles, and the mass of the closed string tachyon is $m_{Tc}^2 = -\frac{4}{\alpha'}$.  This means we have $s + t + u = 4m_{Tc}^2 = -\frac{16}{\alpha'}$ and
\beq
a_c(t) + a_c(s) + a_c(u) = -2 \, .
\eeq
We can use this to rewrite the amplitude in terms of just $s$ and $t$, giving

\beq
\mathcal{A}^4_c(s, t) = 2\pi C \, \frac{\Gamma\left[-\frac{a_c(t)}{2}\right]\Gamma\left[-\frac{a_c(s)}{2}\right]\Gamma\left[1 + \frac{a_c(s)}{2} + \frac{a_c(t)}{2}\right]}{\Gamma\left[-\frac{a_c(s)}{2} - \frac{a_c(t)}{2}\right]\Gamma\left[1 + \frac{a_c(t)}{2}\right]\Gamma\left[1 + \frac{a_c(s)}{2}\right]} \, ,
\eeq
This has t-channel poles whenever $a_c(t)$ is an even integer.  If we expand around one of these poles, we obtain
\beq
\mathcal{A}^4_{c}(s, t)\Bigg|_{a_c(t) = 2n} \approx 2\pi C \, \frac{\Big[\left(1 + \frac{a_c(s)}{2}\right)\left(2 + \frac{a_c(s)}{2}\right)\cdots\left(n+\frac{a_c(s)}{2}\right)\Big]^2}{(n!)^2 \, \left(n - \frac{a_c(t)}{2}\right)} \, .
\eeq
Notice that the leading $s$-dependence of the numerator here is $s^{2n}$.  Usually we expect the exchange of a mediator of spin $J$ to include a factor of $s^{J}$ in it, so here the natural interpretation would be that $2n = J$.  This agrees with the fact that the closed string states correspond only to even spin particles.  In the Regge limit, we obtain
\beq
\mathcal{A}^4_{c}(s, t)\Bigg|_{s \gg t} \approx 2\pi C \, e^{-\frac{i\pi a_c(t)}{2}} \, \left(\frac{\alpha' s}{4}\right)^{a_c(t)} \, \frac{\Gamma\left[-\frac{a_c(t)}{2}\right]}{\Gamma\left[1 + \frac{a_c(t)}{2}\right]} \, .
\eeq
If we suppose we are attempting to Reggeize a closed-string tachyon propagator, this gives
\beq
\frac{1}{t - m_{Tc}^2} \hspace{.25in} \rightarrow \hspace{.25in} -\frac{\alpha'}{4} \, e^{-\frac{i\pi a_c(t)}{2}} \, \left(\frac{\alpha' s}{4}\right)^{a_c(t)} \, \frac{\Gamma\left[-\frac{a_c(t)}{2}\right]}{\Gamma\left[1 + \frac{a_c(t)}{2}\right]} \, .
\eeq

\subsubsection{Modifications for Physical Pomerons and Proton-Proton Scattering}

Suppose instead we want to use a physical pomeron (a Regge trajectory of even spin glueballs), for which the lowest spin state is a spin-2 glueball.  Then we need to start with
\beq
\alpha_g(x) = \alpha_{g0} + \alpha_g'x
\eeq
such that the glueball trajectory has
\beq
J = \alpha_{g0} + \alpha_g'm_J^2
\eeq
We also want to assume we are analyzing elastic proton-proton scattering, so that we should have $s + t + u = 4m_p^2$ and therefore
\beq
\alpha_g(s) + \alpha_g(t) + \alpha_g(u) = 6 + \alpha_g'\left(4m_p^2 - 3m_g^2\right) \equiv \chi_g \, .
\eeq

If we want to modify the Virasoro-Shapiro amplitude in such a way that it has poles on the physical Regge trajectory, we should begin with a ``string inspired'' amplitude, where we replace $a_c(t)$ with $\alpha_g(t) - 2$, so that the first pole corresponds to the spin-2 glueball being exchanged.  We can then rewrite this just in terms of $s$ and $t$, which gives us
\beq
\mathcal{A}_g^4(s, t) = \frac{\Gamma\left[1 - \frac{\alpha_g(t)}{2}\right]\Gamma\left[1 - \frac{\alpha_g(s)}{2}\right]\Gamma\left[1 - \frac{\chi_g}{2} + \frac{\alpha_g(s)}{2} + \frac{\alpha_g(t)}{2}\right]}{\Gamma\left[2 - \frac{\alpha_g(t)}{2} - \frac{\alpha_g(s)}{2}\right]\Gamma\left[2 - \frac{\chi_g}{2} + \frac{\alpha_g(s)}{2}\right]\Gamma\left[2 - \frac{\chi_g}{2} + \frac{\alpha_g(t)}{2}\right]} \, .
\eeq
If we expand this around the $\alpha_g(t) = 2$ pole, we obtain
\beq
\mathcal{A}_g^4\Bigg|_{\alpha_g(t) = 2} \approx \frac{2\pi C}{\Gamma\left[3 - \frac{\chi_g}{2}\right] \, \left(1 - \frac{\alpha_g(t)}{2}\right)} \, ,
\eeq
and on the other hand the Regge limit is
\beq
\mathcal{A}_g^4\Bigg|_{s \gg t} \approx -2\pi C \, e^{-\frac{i\pi\alpha_g(t)}{2}} \, \left(\frac{\alpha_g' s}{2}\right)^{\alpha_g(t) - 2} \, \frac{\Gamma\left[1 - \frac{\alpha_g(t)}{2}\right]}{\Gamma\left[2 - \frac{\chi_g}{2} + \frac{\alpha_g(t)}{2}\right]} \, ,
\eeq
so this leads to the Reggeization procedure
\beq
\frac{1}{t - m_g^2} \hspace{.25in} \rightarrow \hspace{.25in} \frac{\alpha_g'}{2} \, e^{-\frac{i\pi\alpha_g(t)}{2}} \, \left(\frac{\alpha_g' s}{2}\right)^{\alpha_g(t) - 2} \, \frac{\Gamma\left[3 - \frac{\chi_g}{2}\right]\Gamma\left[1 - \frac{\alpha_g(t)}{2}\right]}{\Gamma\left[2 - \frac{\chi_g}{2} + \frac{\alpha_g(t)}{2}\right]} \, .
\eeq

\subsection{Closed 5-String Process}

\subsubsection{In Bosonic String Theory}

The 5-tachyon scattering amplitude in closed bosonic string theory can be written in terms of a double integral, where each integral is over the complex plane, as
\beq
\mathcal{A}_c^{5} = C\int d^2u \, \int d^2v \, |u|^{-a_c(t_1) - 2} \, |v|^{-a_c(t_2) - 2} \, |1 - u|^{-a_c(s_1) - 1} \, |1 - v|^{-a_c(s_2) - 2} \, |1 - uv|^{a_c(s_1) + a_c(s_2) - a_c(s) - 2} \, .
\eeq
In the Regge limit, this integral is dominated by the region where
\beq
u \sim \frac{1}{s_1}, \hspace{1in} v \sim \frac{1}{s_2}
\eeq
which allows us to rewrite it as
\beq
\mathcal{A}_c^{5}\Bigg|_{\mathrm{Regge}} \approx C\int d^2u \, \int d^2 v \, |u|^{-a_c(t_1) - 2} \, |v|^{-a_c(t_2) - 2} \, e^{\frac{\alpha' s_1}{4}(u+\bar{u}) + \frac{\alpha' s_2}{4}(v + \bar{v}) + \frac{\alpha' s}{4}(uv + \bar{u}\bar{v})}
\eeq
We can perform this integral in two separate cases: where $\frac{\alpha'\mu}{4} = \frac{\alpha' s_1 s_2}{4s}$ is large, and where it is small.  In each case we follow methods laid out in the appendix to \cite{Herzog}.

Supposing that $\frac{\alpha' \mu}{4}$ is large, it is necessary to split it up into four pieces, depending on the signs of the real parts of $u$ and $v$, and choose $s_1$ and $s_2$ to have appropriate signs so that each piece separately converges.  We can then perform a change of variables
\beq
w = \pm \frac{\alpha' s_1}{4} \, u, \hspace{1in} z = \pm \frac{\alpha' s_2}{4} \, v
\eeq
for each piece.  This leads to
\beq
\mathcal{A}_c^5\Bigg|_{\mathrm{Regge}} \approx C \, \left[1 + e^{-i\pi a_c(t_1)}\right]\left[1 + e^{-i\pi a_c(t_2)}\right] \, \left(\frac{\alpha' s_1}{4}\right)^{a_c(t_1)}\left(\frac{\alpha' s_2}{4}\right)^{a_c(t_2)}
\eeq
$$
\times \int d^2w \int d^2z \, |w|^{-a_c(t_1) - 2} \, |z|^{-a_c(t_2) - 2} \, e^{-(w + \bar{w}) - (z + \bar{z}) + \frac{4}{\alpha'\mu}(wz + \bar{w}\bar{z})}
$$
We can then expand in $\frac{4}{\alpha'\mu}$ as a small parameter.  The leading term will simply give us
\beq
\mathcal{A}_c^5\Bigg|_{\mathrm{Regge}, \, \frac{\alpha' \mu}{4} \gg 1} \approx C\left\{2\pi e^{-\frac{i\pi a_c(t_1)}{2}} \, \left(\frac{\alpha' s_1}{4}\right)^{a_c(t_1)} \, \frac{\Gamma\left[-\frac{a_c(t_1)}{2}\right]}{\Gamma\left[1 + \frac{a_c(t_1)}{2}\right]}\right\}\left\{2\pi e^{-\frac{i\pi a_c(t_2)}{2}} \, \left(\frac{\alpha' s_2}{4}\right)^{a_c(t_2)} \, \frac{\Gamma\left[-\frac{a_c(t_2)}{2}\right]}{\Gamma\left[1 + \frac{a_c(t_2)}{2}\right]}\right\}
\eeq
which is clearly just two copies of the Regge limit of the 4-string amplitude.

On the other hand, if $\frac{\alpha' \mu}{4}$ is small it makes more sense to use a change of variables that will replace either the integral over $u$ or the integral over $v$ with an integral over $uv$.  In fact, we must do each of these and add the results, because each captures a different saddle point present in the original integral.  That is, we can write
\beq
\mathcal{A}_c^5\Bigg|_{\mathrm{Regge}, \, \frac{\alpha' \mu}{4} \ll 1} \approx \mathcal{I}(s_1, t_1, s_2, t_2) + \mathcal{I}(s_2, t_2, s_1, t_1)
\eeq
with
\beq
\mathcal{I}(s_1, t_1, s_2, t_2) = C \, \int d^2u \, \int d^2v \, |u|^{-a_c(t_1) - 2} \, |v|^{a_c(t_1) - a_c(t_2) - 2} \, e^{\frac{\alpha' s}{4}(u + \bar{u}) + \frac{\alpha' s_2}{4}(v + \bar{v}) + \frac{\alpha' s_1}{4}\left(\frac{u}{v} + \frac{\bar{u}}{\bar{v}}\right)} \, .
\eeq
Then we again have an integral that should be divided into four separate pieces, depending on the signs of the real parts of $u$ and $v$.  In each piece, we choose the signs of $s$ and $s_2$ in order to ensure convergence and perform a change of variables to
\beq
w = \pm \frac{\alpha' s}{4} \, u, \hspace{1in} z = \pm \frac{\alpha' s_2}{4} \, v
\eeq
which gives us
\beq
\mathcal{I}(s_1, t_1, s_2, t_2) = C \, \left[1 + e^{-i\pi a_c(t_1)}\right]\left[1 + e^{i\pi(a_c(t_1) - a_c(t_2))}\right] \, \left(\frac{\alpha' s}{4}\right)^{a_c(t_1)}\left(\frac{\alpha' s_2}{4}\right)^{a_c(t_2) - a_c(t_1)}
\eeq
$$
\times \int d^2w \int d^2z \, |w|^{-a_c(t_1) - 2} \, |z|^{a_c(t_1) - a_c(t_2) - 2} \, e^{-(w + \bar{w}) - (z + \bar{z}) + \frac{\alpha'\mu}{4}\left(\frac{w}{z} + \frac{\bar{w}}{\bar{z}}\right)} \, .
$$
We can then expand this integral in $\frac{\alpha'\mu}{4}$, our small parameter, and the leading term will just give us
\beq
\mathcal{I}(s_1, t_1, s_2, t_2) \approx 4\pi^2C \, e^{-\frac{i\pi a_c(t_2)}{2}} \, \left(\frac{\alpha' s}{4}\right)^{a_c(t_1)}\left(\frac{\alpha' s_2}{4}\right)^{a_c(t_2) - a_c(t_1)} \, \frac{\Gamma\left[-\frac{a_c(t_1)}{2}\right]\Gamma\left[\frac{a_c(t_1)}{2} - \frac{a_c(t_2)}{2}\right]}{\Gamma\left[1 + \frac{a_c(t_1)}{2}\right]\Gamma\left[1 + \frac{a_c(t_2)}{2} - \frac{a_c(t_1)}{2}\right]}
\eeq
When we also use the fact that $s_1, s_2 \approx \sqrt{s\mu}$, we obtain in total
\beq
\mathcal{A}_c^5\Bigg|_{\mathrm{Regge}, \, \frac{\alpha' \mu}{4} \ll 1} \approx 4\pi^2 C \, e^{-\frac{i\pi(a_c(t_1) + a_c(t_2))}{4}} \, \left(\frac{\alpha' s}{4}\right)^{\frac{a_c(t_1) + a_c(t_2)}{2}}
\eeq
$$
\times \left\{\left(\frac{\alpha'\mu}{4}\right)^{\frac{\alpha'(t_2 - t_1)}{4}} \, e^{\frac{i\pi\alpha'(t_1 - t_2)}{8}} \, \frac{\Gamma\left[-\frac{a_c(t_1)}{2}\right]\Gamma\left[\frac{\alpha'(t_1 - t_2)}{4}\right]}{\Gamma\left[1 + \frac{a_c(t_1)}{2}\right]\Gamma\left[1 + \frac{\alpha'(t_2 - t_1)}{4}\right]} \right.
$$
$$
\left. + \left(\frac{\alpha'\mu}{4}\right)^{\frac{\alpha'(t_1 - t_2)}{4}} \, e^{\frac{i\pi\alpha'(t_2 - t_1)}{8}} \, \frac{\Gamma\left[-\frac{a_c(t_2)}{2}\right]\Gamma\left[\frac{\alpha'(t_2 - t_1)}{4}\right]}{\Gamma\left[1 + \frac{a_c(t_2)}{2}\right]\Gamma\left[1 + \frac{\alpha'(t_1 - t_2)}{4}\right]}\right\} \, .
$$
We are more interested in this approximation, because in the physical process we live in the ``small $\mu$'' regime.  In principle to follow the same procedure we did for the 4-string process, we should now compare this Regge limit to a pole expansion.  This is somewhat difficult because we don't have the exact amplitude in closed form.  Instead, we will assume that the ``large $\mu$'' regime indeed just gives us two copies of the Reggeization procedure for the 4-string process, and use this to deduce the appropriate procedure for the ``small $\mu$'' regime.  This gives
\beq
\frac{1}{(t_1 - m_{Tc}^2)(t_2 - m_{Tc}^2)} \hspace{.25in} \rightarrow \hspace{.25in} \left(\frac{4}{\alpha'}\right)^2\, e^{-\frac{i\pi(a_c(t_1) + a_c(t_2))}{4}} \, \left(\frac{\alpha' s}{4}\right)^{\frac{a_c(t_1) + a_c(t_2)}{2}}
\eeq
$$
\times \left\{\left(\frac{\alpha'\mu}{4}\right)^{\frac{\alpha'(t_2 - t_1)}{4}} \, e^{\frac{i\pi\alpha'(t_1 - t_2)}{8}} \, \frac{\Gamma\left[-\frac{a_c(t_1)}{2}\right]\Gamma\left[\frac{\alpha'(t_1 - t_2)}{4}\right]}{\Gamma\left[1 + \frac{a_c(t_1)}{2}\right]\Gamma\left[1 + \frac{\alpha'(t_2 - t_1)}{4}\right]} \right.
$$
$$
\left. + \left(\frac{\alpha'\mu}{4}\right)^{\frac{\alpha'(t_1 - t_2)}{4}} \, e^{\frac{i\pi\alpha'(t_2 - t_1)}{8}} \, \frac{\Gamma\left[-\frac{a_c(t_2)}{2}\right]\Gamma\left[\frac{\alpha'(t_2 - t_1)}{4}\right]}{\Gamma\left[1 + \frac{a_c(t_2)}{2}\right]\Gamma\left[1 + \frac{\alpha'(t_1 - t_2)}{4}\right]}\right\} \, .
$$

\subsubsection{Modifications for Physical Pomerons and Proton-Proton Scattering}

Now we need to establish how to modify the Reggeization procedure to account for the physical Pomeron trajectory and the masses of the protons.  To do this properly we ought to begin by writing the exact amplitude in a way that is symmetric in all five external particle momenta, and then use the appropriate mass-shell conditions to rewrite this in terms of $\{s, s_1, s_2, t_1, t_2\}$ and the masses of the particles.  Our experience with the 4-string process suggests that in doing so we would introduce a dependence on factors such as the $\chi_g$ we saw earlier.  Since we do not have the amplitude in closed form, we will instead simply propose the following modification, which ensures the correct pole structure and is based on what we saw for the 4-string amplitude:

\beq
\frac{1}{(t_1 - m_g^2)(t_2 - m_g^2)} \hspace{.25in} \rightarrow \hspace{.25in} -\frac{1}{s^2} \, \Gamma\left[3 - \frac{\chi_g}{2}\right]^2 \, e^{-\frac{i\pi(\alpha_g(t_1) + \alpha_g(t_2))}{4}} \, \left(\frac{\alpha_g' s}{2}\right)^{\frac{\alpha_g(t_1) + \alpha_g(t_2)}{2}}
\eeq
$$
\times  \left\{\left(\frac{\alpha_g'\mu}{2}\right)^{\frac{\alpha_g'(t_2 - t_1)}{2}} \, e^{\frac{i\pi\alpha_g'(t_1 - t_2)}{4}} \, \frac{\Gamma\left[1 - \frac{\alpha_g(t_1)}{2}\right]\Gamma\left[\frac{\alpha'_g(t_1 - t_2)}{2}\right]}{\Gamma\left[2 - \frac{\chi_g}{2} + \frac{\alpha_g(t_1)}{2}\right]\Gamma\left[3 - \frac{\chi_g}{2} + \frac{\alpha_g'(t_2 - t_1)}{2}\right]} \right.
$$
$$
\left. +  \left(\frac{\alpha_g'\mu}{2}\right)^{\frac{\alpha_g'(t_1 - t_2)}{2}} \, e^{\frac{i\pi\alpha_g'(t_2 - t_1)}{4}} \, \frac{\Gamma\left[1 - \frac{\alpha_g(t_2)}{2}\right]\Gamma\left[\frac{\alpha'_g(t_2 - t_1)}{2}\right]}{\Gamma\left[2 - \frac{\chi_g}{2} + \frac{\alpha_g(t_2)}{2}\right]\Gamma\left[3 - \frac{\chi_g}{2} + \frac{\alpha_g'(t_1 - t_2)}{2}\right]} \right\}
$$

\subsection{Open 5-String Process}

\subsubsection{In Bosonic String Theory}

The 5-tachyon scattering amplitude in open bosonic string theory can be written as the double integral
\beq
\mathcal{A}_o^5 = 2iC\int_{-\infty}^{\infty} \int_{-\infty}^{\infty} dx \, dy \, |x|^{-1 - a_o(t_1)}|y|^{-1 - a_o(t_2)}|1 - x|^{-1 - a_o(s_1)}|1 - y|^{-1 - a_o(s_2)}|1 - xy|^{a_o(s_1) + a_o(s_2) - a_o(s)} \, .
\eeq
In the Regge limit, these integrals should be dominated by regions near the four points in the $xy$-plane given by
\beq
x \sim \pm \frac{1}{s_1}, \hspace{1in} y \sim \pm \frac{1}{s_2} \, .
\eeq
We can split up the regions of integration to produce four terms of the form
\beq
\tilde{\mathcal{A}}_{o}^{\pm, \pm} = 2iC\int_{\Re^{\pm}} dx \int_{\Re^{\pm}} dy \, |x|^{-1 - a_o(t_1)}|y|^{-1 - a_o(t_2)}|1 - x|^{-1 - a_o(s_1)}|1 - y|^{-1 - a_o(s_2)}|1 - xy|^{a_o(s_1) + a_o(s_2) - a_o(s)} \, ,
\eeq
each of which contains one of these saddle points.  These four terms are analogous to the terms $\tilde{\mathcal{A}}_o(s, t)$ and $\tilde{\mathcal{A}}_o(u, t)$ we have in the 4-string case, and we will want to take a linear combination of them in order to project out poles associated with the exchange of even spin particles.  Specifically, the combination
\beq
\mathcal{A}_{o, \mathrm{odd}}^{5} = \tilde{\mathcal{A}}_{o}^{+, +} - \tilde{\mathcal{A}}_{o}^{-, +} - \tilde{\mathcal{A}}_{o}^{+, -} + \tilde{\mathcal{A}}_{o}^{-, -}
\eeq
will work (as will become apparent shortly.)  With an appropriate change of variables, we can rewrite them as
\beq
\tilde{\mathcal{A}}_{o}^{\pm, \pm} = 2iC\int_{0}^{\infty} dx \int_{0}^{\infty} dy \, x^{-1 - a_o(t_1)} \, y^{-1 - a_o(t_2)} \, (1 \mp x)^{-1 - a_o(s_1)} \, (1 \mp y)^{-1 - a_o(s_2)} \, (1 - (\pm)(\pm)xy)^{a_o(s_1) + a_o(s_2) - a_o(s)} \, .
\eeq
Working in the Regge limit near $x \sim \frac{1}{s_1}$ and $y \sim \frac{1}{s_2}$ then gives us
\beq
\tilde{\mathcal{A}}_{o}^{\pm, \pm}\Bigg|_{\mathrm{Regge}} \approx 2iC\int_{0}^{\infty} dx \int_{0}^{\infty} dy \, x^{-1 - a_o(t_1)} \, y^{-1 - a_o(t_2)} \, e^{\pm\alpha's_1x \pm \alpha' s_2y + (\pm)(\pm)\alpha' s xy} \, .
\eeq

Again, at this point our treatment will be different depending on whether $\alpha'\mu$ is large, or small.  If it is large, then we should proceed by performing the change of variables
\beq
w = \mp \alpha' s_1 x, \hspace{1in} z = \mp \alpha' s_2 y
\eeq
and obtain
\beq
\tilde{\mathcal{A}}_{o}^{+,+}\Bigg|_{\mathrm{Regge}} \approx 2iC \, e^{-i\pi a_o(t_1)} \, e^{-i\pi a_o(t_2)} \, (\alpha' s_1)^{a_o(t_1)} \, (\alpha' s_2)^{a_o(t_2)} \, \mathcal{B} \, ,
\eeq
\beq
\tilde{\mathcal{A}}_{o}^{-,+}\Bigg|_{\mathrm{Regge}} \approx 2iC \, e^{-i\pi a_o(t_2)} \, (\alpha' s_1)^{a_o(t_1)} \, (\alpha' s_2)^{a_o(t_2)} \, \, \mathcal{B} \, ,
\eeq
\beq
\tilde{\mathcal{A}}_{o}^{+,-}\Bigg|_{\mathrm{Regge}} \approx 2iC \, e^{-i\pi a_o(t_1)} \, (\alpha' s_1)^{a_o(t_1)} \, (\alpha' s_2)^{a_o(t_2)} \, \, \mathcal{B} \, ,
\eeq
\beq
\tilde{\mathcal{A}}_{o}^{-,-}\Bigg|_{\mathrm{Regge}} \approx 2iC \, (\alpha' s_1)^{a_o(t_1)} \, (\alpha' s_2)^{a_o(t_2)} \, \, \mathcal{B} \, ,
\eeq
with
\beq
\mathcal{B} = \int_0^{\infty} dz \int_0^{\infty} dw \, w^{-1 - a_o(t_1)} \, z^{-1 - a_0(t_2)} \, e^{-w-z + \frac{zw}{\alpha'\mu}} \, .
\eeq
If we then expand in the small parameter $\frac{1}{\alpha'\mu}$, the leading term will just give us
\beq
\mathcal{B}\Bigg|_{\alpha'\mu \gg 1} \approx \Gamma[-a_o(t_1)]\Gamma[-a_o(t_2)] \, .
\eeq
Then we have
\beq
\mathcal{A}_{o, \mathrm{odd}}^{5}\Bigg|_{\mathrm{Regge}, \, \alpha'\mu \gg 1} \approx -8iC \, e^{-\frac{i\pi a_o(t_1)}{2}} \, e^{-\frac{i\pi a_o(t_2)}{2}} \, \sin\left[\frac{\pi a_o(t_1)}{2}\right] \, \sin\left[\frac{\pi a_o(t_2)}{2}\right] \, (\alpha' s_1)^{a_o(t_1)} \, (\alpha' s_2)^{a_o(t_2)} \, \Gamma[-a_o(t_1)]\Gamma[-a_o(t_2)]
\eeq
Note that at this stage it is very clear the above linear combination will have no poles for $a_o(t_i)$ equal to an even integer, and (just as we did for the closed 5-string amplitude when $\mu$ was large), we are essentially obtaining two copies of the 4-string result.

On the other hand, if we have $\alpha'\mu$ small, we should replace either the integral over $x$ or the integral over $y$ with an integral over $xy$, and in fact each of these captures the effect of a different saddle point, so we need to do both and add them together, giving
\beq
\tilde{\mathcal{A}}^{\pm, \pm}_{o}\Bigg|_{\mathrm{Regge}, \alpha'\mu \ll 1} \approx \mathcal{I}^{\pm, \pm}(s_1, t_1, s_2, t_2) + \mathcal{I}^{\pm, \pm}(s_2, t_2, s_1, t_1) \, ,
\eeq
with
\begin{eqnarray}
\mathcal{I}^{+,+}(s_1, t_1, s_2, t_2) & = & 2iC\int_0^\infty\int_0^\infty dx \, dy \, x^{-1 - a_o(t_1)} \, y^{-1 - a_o(t_2) + a_o(t_1)} \, e^{\alpha' s x + \alpha' s_2 y + \frac{\alpha' s_1 x}{y}} \, , \\
\mathcal{I}^{-,+}(s_1, t_1, s_2, t_2) & = & 2iC\int_0^\infty\int_0^\infty dx \, dy \, x^{-1 - a_o(t_1)} \, y^{-1 - a_o(t_2) + a_o(t_1)} \, e^{-\alpha' s x + \alpha' s_2 y - \frac{\alpha' s_1 x}{y}} \, , \nonumber \\
\mathcal{I}^{+,-}(s_1, t_1, s_2, t_2) & = & 2iC\int_0^\infty\int_0^\infty dx \, dy \, x^{-1 - a_o(t_1)} \, y^{-1 - a_o(t_2) + a_o(t_1)} \, e^{-\alpha' s x - \alpha' s_2 y + \frac{\alpha' s_1 x}{y}} \, , \nonumber \\
\mathcal{I}^{-,-}(s_1, t_1, s_2, t_2) & = & 2iC\int_0^\infty\int_0^\infty dx \, dy \, x^{-1 - a_o(t_1)} \, y^{-1 - a_o(t_2) + a_o(t_1)} \, e^{\alpha' s x - \alpha' s_2 y - \frac{\alpha' s_1 x}{y}} \, . \nonumber
\end{eqnarray}
We can then perform a change of variables of the form
\beq
w = \pm \alpha' s x, \hspace{1in} z = \pm \alpha' s_2 y
\eeq
for each case, giving
\begin{eqnarray}
\mathcal{I}^{+,+}(s_1, t_1, s_2, t_2) & = & 2iC \, e^{-i\pi a_o(t_2)} \, (\alpha' s)^{a_o(t_1)} \, (\alpha' s_2)^{a_o(t_2) - a_o(t_1)} \, \mathcal{B}_{12} \, , \\
\mathcal{I}^{-,+}(s_1, t_1, s_2, t_2) & = & 2iC \, e^{-i\pi a_o(t_2)} \, e^{i\pi a_o(t_1)} \, (\alpha' s)^{a_o(t_1)} \, (\alpha' s_2)^{a_o(t_2) - a_o(t_1)} \, \mathcal{B}_{12} \, , \nonumber \\
\mathcal{I}^{+,-}(s_1, t_1, s_2, t_2) & = & 2iC  \, (\alpha' s)^{a_o(t_1)} \, (\alpha' s_2)^{a_o(t_2) - a_o(t_1)} \, \mathcal{B}_{12} \, , \nonumber \\
\mathcal{I}^{-,-}(s_1, t_1, s_2, t_2) & = & 2iC \, e^{-i\pi a_o(t_1)} \, (\alpha' s)^{a_o(t_1)} \, (\alpha' s_2)^{a_o(t_2) - a_o(t_1)} \, \mathcal{B}_{12} \, , \nonumber
\end{eqnarray}
with
\beq
\mathcal{B}_{12} = \int_0^{\infty}\int_{0}^{\infty} dw \, dz \, w^{-1 - a_o(t_1)} \, z^{-1 - a_o(t_2) + a_o(t_1)} \, e^{-z - w + \frac{\alpha' \mu z}{w}} \, .
\eeq
We can expand this in the small parameter $\alpha'\mu$, and the leading term just gives
\beq
\mathcal{B}_{12}\Bigg|_{\alpha'\mu \ll 1} \approx \Gamma[-a_o(t_1)]\Gamma[a_o(t_1) - a_o(t_2)]
\eeq
so that
\beq
\mathcal{A}^5_{o, \, \mathrm{odd}}\Bigg|_{\mathrm{Regge}, \, \alpha'\mu \ll 1} \approx 8C \, e^{-\frac{i\pi(a_o(t_1) + a_o(t_2)}{4}} \, \cos\left[\frac{\pi\alpha'(t_2 - t_1)}{2}\right] \, (\alpha' s)^{\frac{a_o(t_1) + a_o(t_2)}{2}}
\eeq
$$
\times \left\{e^{\frac{i\pi\alpha'(t_1 - t_2)}{4}} \, (\alpha'\mu)^{\frac{\alpha'(t_2 - t_1)}{2}} \, \sin\left[\frac{\pi a_o(t_1)}{2}\right] \, \Gamma[-a_o(t_1)] \, \Gamma[\alpha'(t_1 - t_2)] \right.
$$
$$
+ \left. e^{\frac{i\pi\alpha'(t_2 - t_1)}{4}} \, (\alpha'\mu)^{\frac{\alpha'(t_1 - t_2)}{2}} \, \sin\left[\frac{\pi a_o(t_2)}{2}\right] \, \Gamma[-a_o(t_2)] \, \Gamma[\alpha'(t_2 - t_1)] \right\} \, .
$$
As with the 5-closed-string case, we can now work out what the appropriate Reggeization procedure must be when $\alpha'\mu$ is small by assuming that the Reggeization for large $\alpha'\mu$ is just two copies of the 4-string procedure.  This gives
\beq
\frac{1}{(t_1 - m_1^2)(t_2 - m_1^2)} \hspace{.25in} \rightarrow \hspace{.25in} i\alpha'^2 e^{-\frac{i\pi(a_o(t_1) + a_o(t_2)}{4}} \, \cos\left[\frac{\pi\alpha'(t_2 - t_1)}{2}\right] \, (\alpha' s)^{\frac{a_o(t_1) + a_o(t_2)}{2} - 1}
\eeq
$$
\times \left\{e^{\frac{i\pi\alpha'(t_1 - t_2)}{4}} \, (\alpha'\mu)^{\frac{\alpha'(t_2 - t_1)}{2} - 1} \, \sin\left[\frac{\pi a_o(t_1)}{2}\right] \, \Gamma[-a_o(t_1)] \, \Gamma[\alpha'(t_1 - t_2)] \right.
$$
$$
+ \left. e^{\frac{i\pi\alpha'(t_2 - t_1)}{4}} \, (\alpha'\mu)^{\frac{\alpha'(t_1 - t_2)}{2} -1} \, \sin\left[\frac{\pi a_o(t_2)}{2}\right] \, \Gamma[-a_o(t_2)] \, \Gamma[\alpha'(t_2 - t_1)] \right\} \, .
$$

\subsubsection{Modifications for Physical Pomerons and Proton-Proton Scattering}

Finally, we need to incorporate the modifications to this Reggeization procedure that follow from assuming we are looking at a physical trajectory of mesons (either the $\rho$ or the $\omega$ trajectory) and that the incoming and outgoing particles are protons and an $\eta$ meson.  However, based on what we saw for the 4-string case, it is reasonable to assume that the only change should be replacing the open string Regge trajectory $a_o(x)$ with the physical Regge trajectory $\alpha_v(x)$, which leads to

\beq
\frac{1}{(t_1 - m_v^2)(t_2 - m_v^2)} \hspace{.25in} \rightarrow \hspace{.25in} i\frac{\alpha_v'}{s} \, e^{-\frac{i\pi(\alpha_v(t_1) + \alpha_v(t_2)}{4}} \, \cos\left[\frac{\pi\alpha'_v(t_2 - t_1)}{2}\right] \, (\alpha_v' s)^{\frac{\alpha_v(t_1) + \alpha_v(t_2)}{2}}
\eeq
$$
\times \left\{e^{\frac{i\pi\alpha_v'(t_1 - t_2)}{4}} \, (\alpha_v'\mu)^{\frac{\alpha_v'(t_2 - t_1)}{2} - 1} \, \sin\left[\frac{\pi \alpha_v(t_1)}{2}\right] \, \Gamma[-\alpha_v(t_1)] \, \Gamma[\alpha'_v(t_1 - t_2)] \right.
$$
$$
+ \left. e^{\frac{i\pi\alpha_v'(t_2 - t_1)}{4}} \, (\alpha_v'\mu)^{\frac{\alpha_v'(t_1 - t_2)}{2} - 1} \, \sin\left[\frac{\pi \alpha_v(t_2)}{2}\right] \, \Gamma[-\alpha_v(t_2)] \, \Gamma[\alpha_v'(t_2 - t_1)] \right\} \, .
$$

\end{appendix}

\end{document}